\documentclass[twocolumn,showpacs,amsmath,preprintnumbers,pre]{revtex4}

\usepackage{graphicx}
\usepackage{amsmath}
\usepackage{amssymb}
\usepackage{natbib}

\begin{document}

\title{Structural signatures of the unjamming transition at zero temperature}

\author{Leonardo E. Silbert}

\affiliation{James Franck Institute, University of Chicago,
   Chicago, IL 60637}

\author{Andrea J. Liu}

\affiliation{Department of Physics and Astronomy, University of Pennsylvania,
  PA 19104}

\author{Sidney R. Nagel}

\affiliation{The James Franck Institute, University of Chicago, Chicago, IL 60637}

\date{\today}

\begin{abstract}

  We study the pair correlation function $g(r)$ for zero-temperature,
  disordered, soft-sphere packings just above the onset of jamming. We find
  distinct signatures of the transition in both the first and split second
  peaks of this function.  As the transition is approached from the jammed
  side (at higher packing fraction) the first peak diverges and narrows on the
  small-$r$ side to a delta-function.  On the high-$r$ side of this peak,
  $g(r)$ decays as a power-law. In the split second peak, the two subpeaks are
  both singular at the transition, with power-law behavior on their low-$r$
  sides and step-function drop-offs on their high-$r$ sides.  These
  singularities at the transition are reminiscent of empirical criteria that
  have previously been used to distinguish glassy structures from liquid ones.

\end{abstract}

\maketitle

It is only natural to suspect that the dramatic dynamical arrest that occurs
as a liquid is cooled into a glass must be accompanied by a signature in the
underlying atomic arrangements. However, the atomic configurations in the
liquid and glass are strikingly similar to one another. Over the years, the
challenge to identify a subtle structural difference between the two states
has led to the proposal of several empirical criteria
\cite{cargill1,abraham1,hiwatari1}. In this paper, we revisit an old idea, due
to Bernal \cite{bernal5}, of using static sphere packings to gain insight into
the structure of amorphous systems
\cite{scott2,cargill2,bennett1,finney2,finney4,torquato3}. We find that, with
decreasing density, the structure of such packings changes distinctly as they
unjam.

It is particularly revealing to study static packings of spheres interacting
via a finite-ranged, purely-repulsive potential. There is a fundamental change
in the mechanical properties in such systems, reminiscent of the glass
transition \cite{liu1}, as the packing fraction $\phi$ is varied across a
well-defined unjamming/jamming transition at $\phi_{c}$, which was found to
coincide with the value of random close packing $\approx 0.64$
\cite{ohern2,ohern3}. Above $\phi_{c}$ the system has nonzero static shear and
bulk moduli, while below $\phi_{c}$ it costs no energy to shear or compress
the system by an infinitesimal amount. Moreover, the unjamming transition in
many ways resembles a critical point
\cite{weaire1,durian5,ohern2,ohern3,makse1,leo14,wyart1,jen1,henkes1}, with
many quantities, including a diverging length scale, behaving as power-laws as
the transition is approached from higher density. Here we show that the
unjamming transition exhibits clearly identifiable structural signatures
associated with diverging quantities, even though both the jammed and the
unjammed states are disordered. These structural characteristics are echoed in
some of the empirical criteria \cite{cargill1,abraham1,hiwatari1} that have
previously been proposed for the glass transition.

The simulations reported here are for monodisperse, soft spheres of diameter
$\sigma$ that interact through the potential $V(r) =
(V_{0}/\alpha)(1-r)^{\alpha}$, for $ r<1$, and $V(r)=0$ when $r \geq 1$.
Here, $r$ is the center-to-center separation between two particles, measured
in units of $\sigma$. We have studied both the harmonic, $\alpha = 2$, and the
Hertzian, $\alpha = 2.5$, cases.  Particles are defined to be in contact if
they overlap. Our three-dimensional systems consist of $1024 \le N \le 10000$
spheres in periodic, cubic, simulation cells. To enable a systematic study of
the approach to the unjamming transition, we employ conjugate-gradient energy
minimization \cite{press1} to obtain $T = 0$ configurations at various packing
fractions $\phi$. We average over ensembles of configurations at the same
distance from the transition point, {\it i.e.,} at the same values of
$\Delta\phi \equiv \phi-\phi_{c}$, which is equivalent to averaging over
systems with the same pressure \cite{ohern2,ohern3}.

\begin{figure}[h]
  \includegraphics[width=7cm]{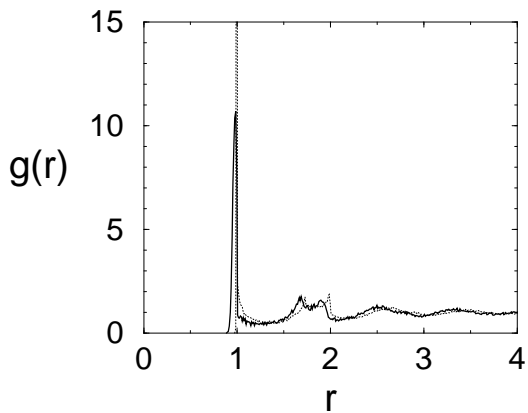}
  \caption  {The pair correlation function $g(r)$ vs. $r$ at two extreme values of
    $\Delta \phi \equiv \phi-\phi_{c}=10^{-6}$ (dotted line) and $10^{-1}$
    (solid line). The maximum value of the first peak height is higher and its
    width narrower for the lower value of $\Delta\phi$. For $\Delta\phi =
    10^{-6}$, the first peak maximum is approximately $10^6$, far beyond the
    scale of the graph.}
 \label{fig1}
\end{figure}

Structural signatures of jamming are more evident \cite{ohern3} in the pair
correlation function, $g(r)$, than in the structure factor, $S(k)$, even
though the two functions are simply related by a Fourier transform. We
therefore focus on the structural characteristics that signal the approach of
the zero-temperature transition from the jammed side, through a detailed
analysis of $g(r)$ computed with extraordinary resolution. In particular, we
study two features of $g(r)$ that undergo dramatic changes at the unjamming
transition: the first peak and the split second peak ({\it i.~e.}, the two
subpeaks that merge with increasing temperature to become the second peak in a
typical dense liquid). A plot of $g(r)$ is shown in Fig.~\ref{fig1} above the
jamming transition.

\section {First peak of $g(r)$}
The dominant feature in $g(r)$ is the first tall, sharp peak at $r_{\rm peak}$
(Fig.~\ref{fig1}). Precisely at the jamming threshold, $\Delta\phi \equiv
\phi-\phi_{c} =0$, this peak is a $\delta$-function at $r_{\rm peak}=1$;
$g(r)$ is precisely zero for $r<1$ and has a power-law tail extending to $r >
1$. The weight under the $\delta$-function is the coordination number at
contact, $Z_{\rm contact}$. (As we discuss below, $Z_{\rm contact}$ is a few
percent less than the isostatic coordination number, $Z_{c} = 2d = 6$ for our
$d=3$-dimensional systems \cite{alexander1,ohern2,ohern3}.) For $\Delta\phi
>0$, there is some overlap between particles so that the delta-function peak
broadens and shifts to $r_{\rm peak} < 1$. The broadening produces a tail
extending to $r < r_{\rm peak}$ that disappears in the limit where
$\Delta\phi$ vanishes.

\begin{figure}[h]
  \includegraphics[width=7cm]{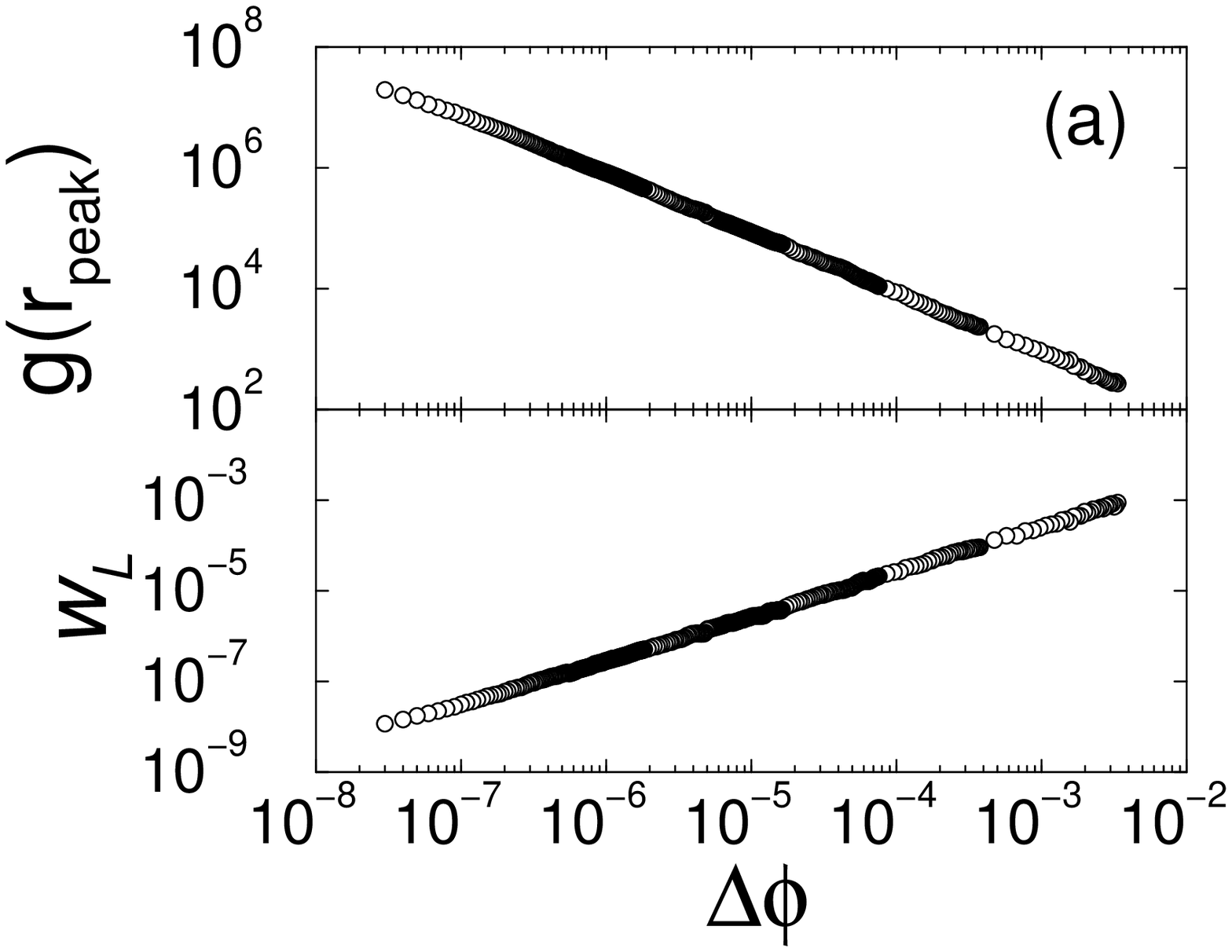}
  \includegraphics[width=7cm]{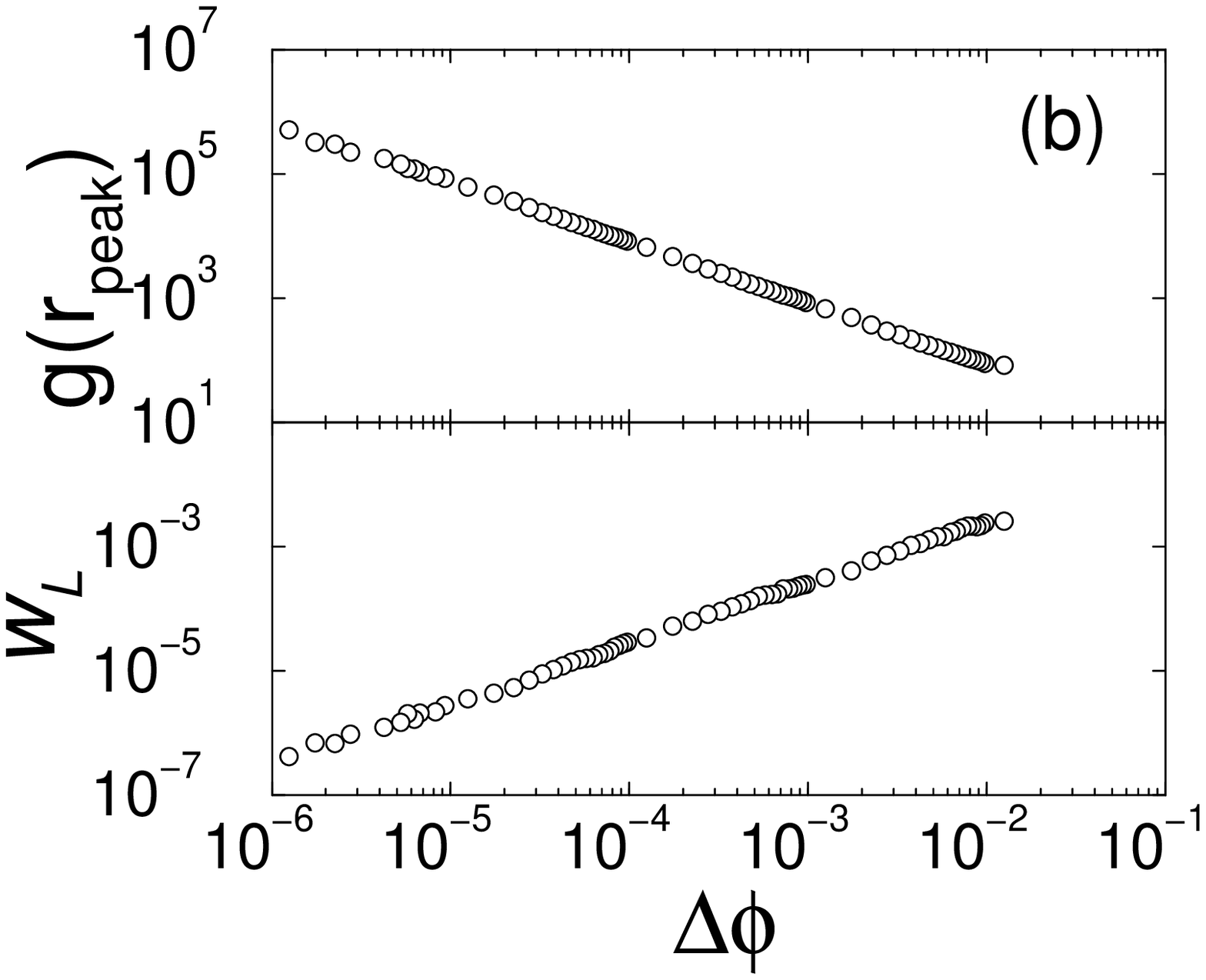}
  \caption{The height, $g(r_{\rm peak})$, (top panel) and the left-hand width, $w_{\rm
      L}$, (bottom panel) of the nearest-neighbor peak of $g(r)$, over several
    orders of magnitude of $\Delta\phi$, for monodisperse spheres with purely
    repulsive (a) harmonic spring and (b) Hertzian interactions.}
  \label{fig2}
\end{figure}

By analyzing the height of the first peak and its left-hand width, we showed
\cite{ohern3} that the peak approaches a delta-function as $\Delta \phi$
decreases towards zero. We have since obtained more systems over a wider range
of $\Delta\phi$. Figure \ref{fig2} shows the dependence on $\Delta\phi$ of the
first-peak height $g(r_{\rm peak})$ and its left-hand width $w_{\rm L}$ for
both harmonic and Hertzian potentials. Independent of the interaction
potential, we find:
\begin{eqnarray}
    g(r_{\rm peak})& \sim & \Delta\phi^{-1.0}\\
    \label{eqn1}
    w_{\rm L} & \sim & \Delta\phi^{1.0}
    \label{eqn2}
\end{eqnarray}
These scalings are consistent with the area of the peak approaching a constant
in the limit $\Delta\phi \rightarrow 0$.

\begin{figure}[h]
  \includegraphics[width=3.5cm]{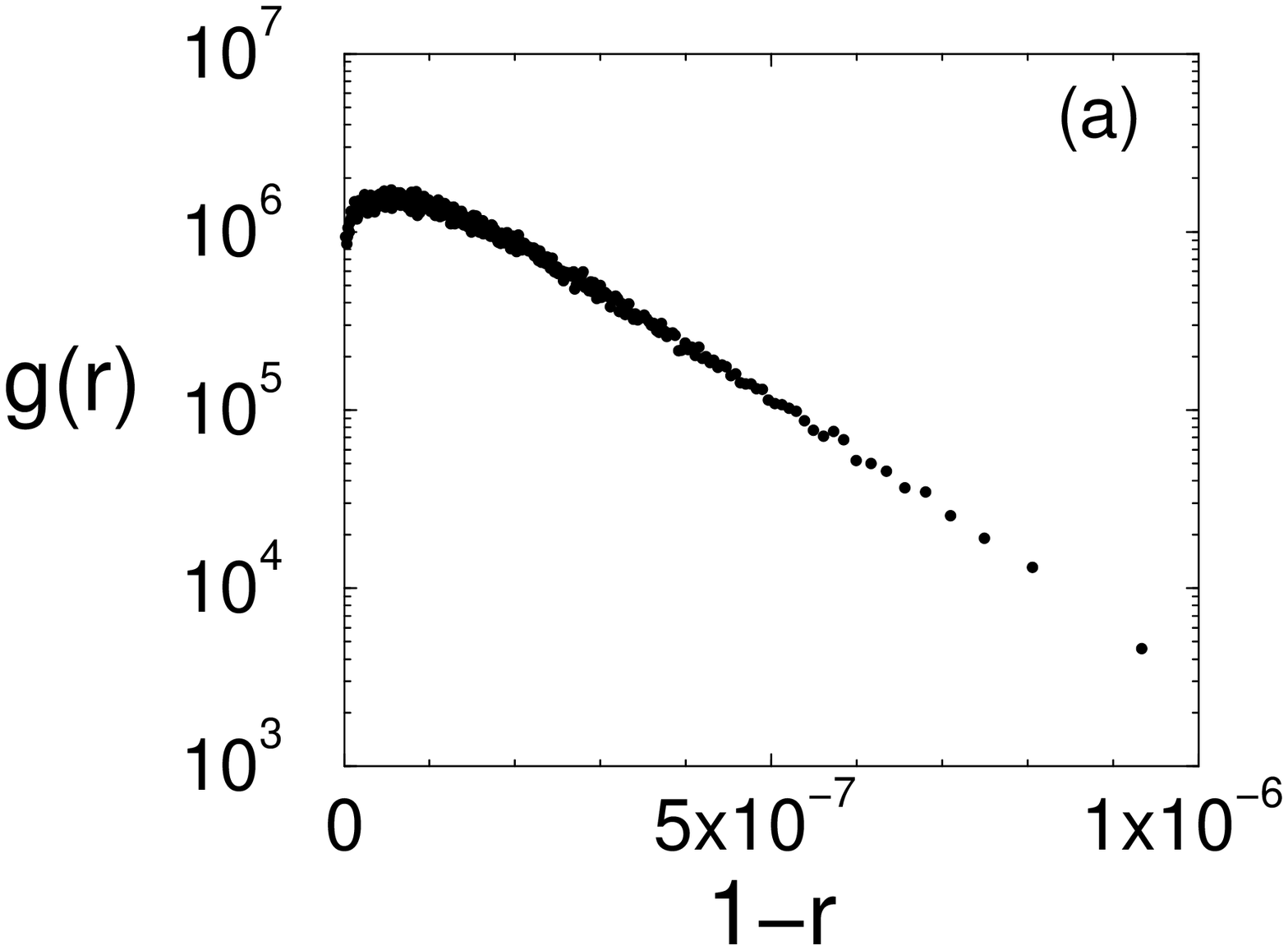}\hfil
  \includegraphics[width=3.5cm]{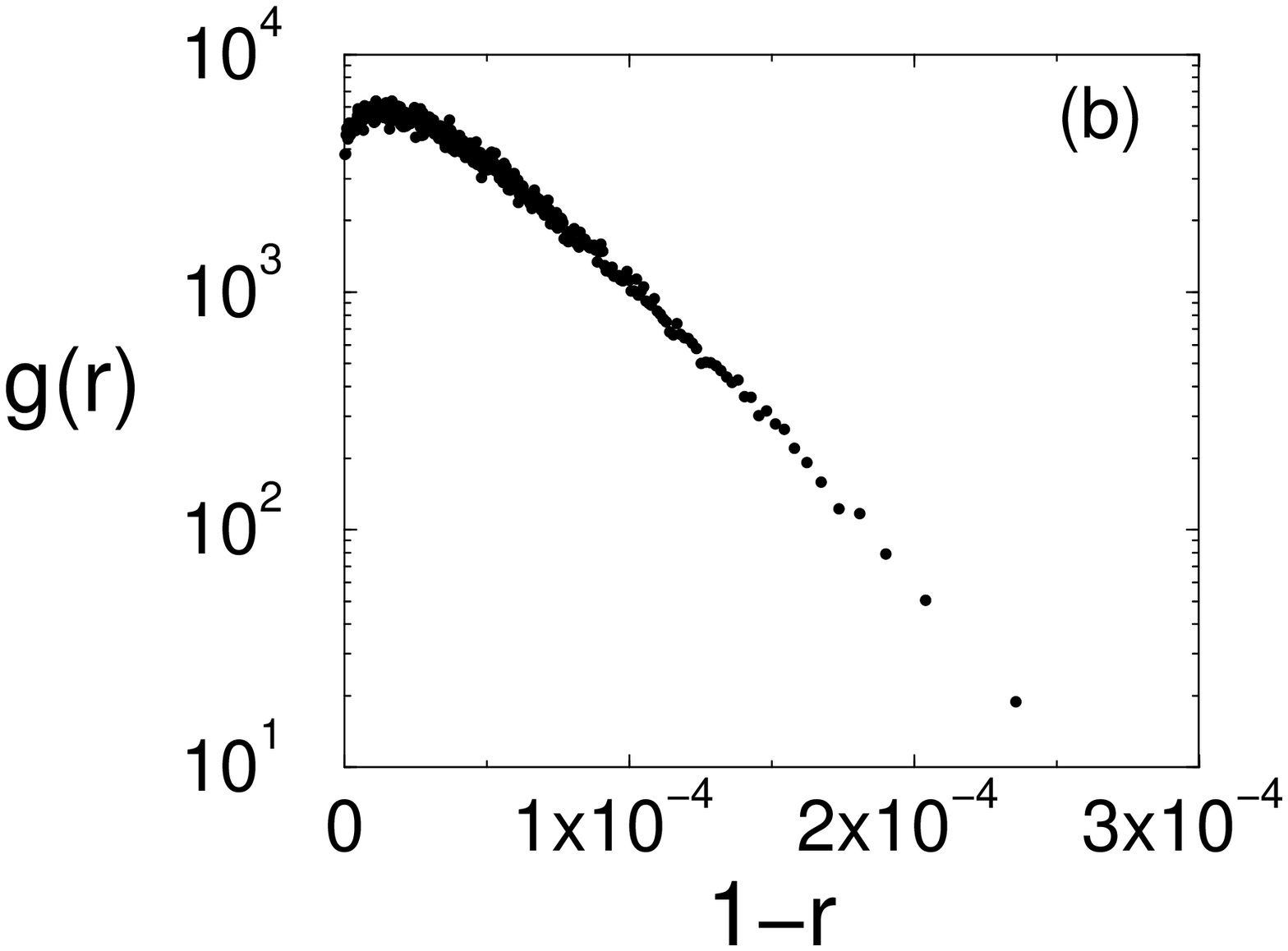}\hfil
  \includegraphics[width=3.25cm]{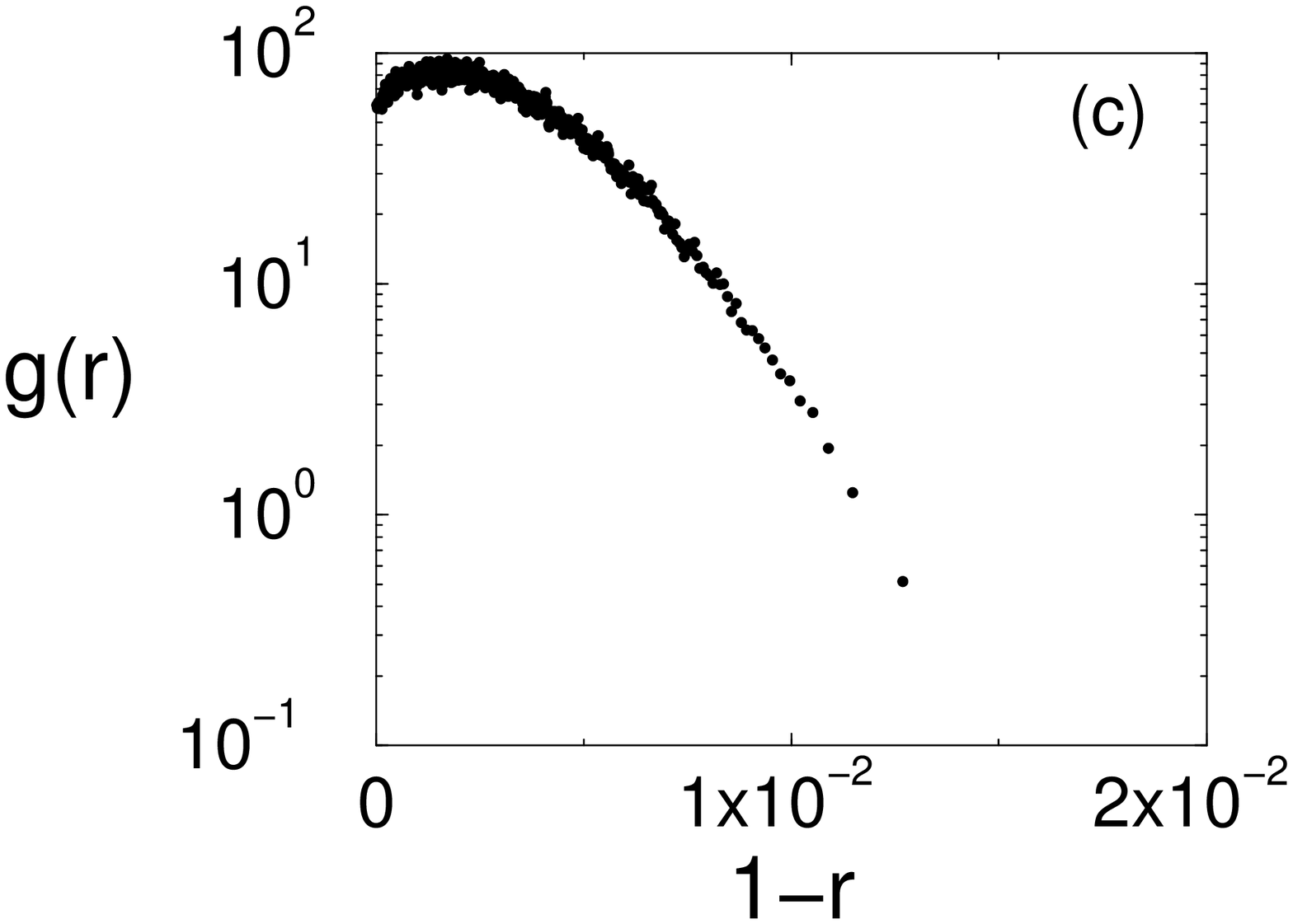}\hfil
  \includegraphics[width=3.25cm]{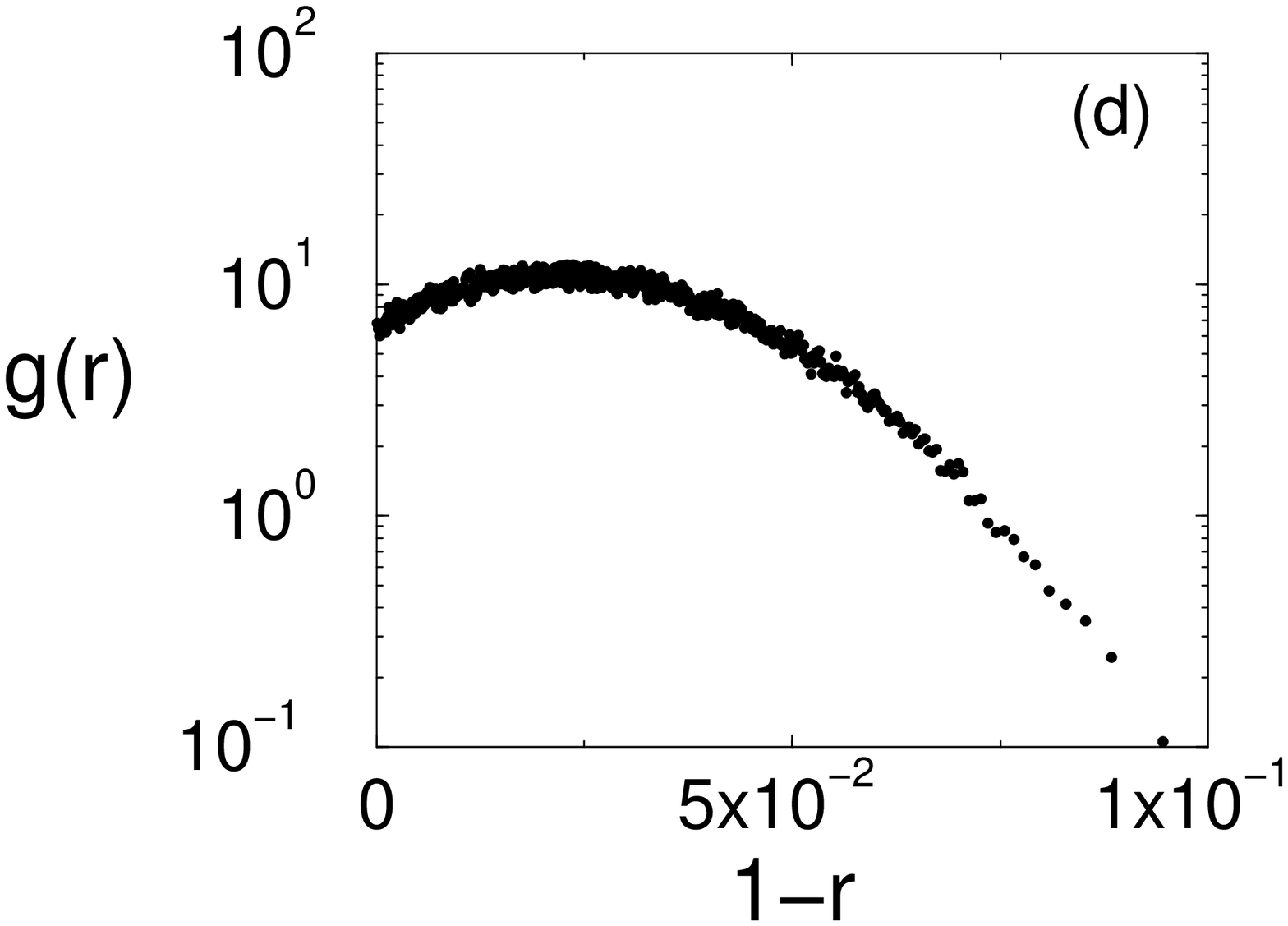}\hfil
  \caption{Pair distribution function $g(r)$ up to contact, $0 < r <
    1$, on a linear-$\log$ scale, for $\Delta \phi =$ (a) $10^{-6}$, (b)
    $10^{-4}$, (c) $10^{-2}$, and (d) $10^{-1}$.}
  \label{fig3}
\end{figure}

We turn now to the shape of the first peak in $g(r)$ at $r<1$ for a system at
$\Delta\phi = 1\times 10^{-6}$, just above the jamming transition. As shown in
Fig.~\ref{fig3}(a), for the region $r<1$, $g(r)$ is almost strictly
exponential with only a small curvature near its peak. It can be fit with the
functional form:
\begin{equation}
  g(r<\sigma) = g_{\circ}\exp \left (-\left [\frac{\alpha_{1}}{\delta} +
      \frac{\alpha_{2}^{2}}{\delta^{2}} \right ] ^{-1}\right )
  \label{eqn3}
\end{equation}
where $\delta\equiv 1-r/r_{\rm peak}$, with $\alpha_{1}=1.4 \times 10^{-7}$
and $\alpha_{2}=1.2 \times 10^{-7}$.

\begin{figure}[h]
  \includegraphics[width=7cm]{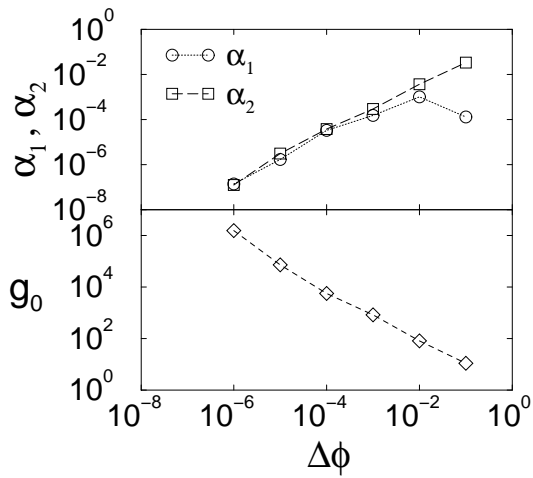}
  \caption{Evolution of the parameters $\alpha_{1}$ and $\alpha_{2}$
    (top) and $g_{\circ}$ (bottom) in Eq.~\ref{eqn3}, with $\Delta\phi$, for
    harmonic repulsions. The parameters $\alpha_{1,2}$ are clearly related to
    the pressure, while $g_{\circ}$ is associated with $g(r_{peak})$.}
  \label{fig4}
\end{figure}

As we compress the system above $\phi_{c}$ we see that the exponential
behavior of the tail at $r<1$ gradually becomes more Gaussian as the system is
compressed above the transition. This is shown in Fig.~\ref{fig3}a-d. We can
still use Eq.~\ref{eqn3} to fit the shape, however, with different
coefficients $\alpha_{1}$ and $\alpha_{2}$. In Fig.~\ref{fig4} we show the
evolution of $\alpha_{1}$ and $\alpha_{2}$ with $\Delta\phi$. This evolution
occurs more rapidly for Hertzian (not shown) than for harmonic spheres.

\begin{figure}[h]
  \includegraphics[width=7cm]{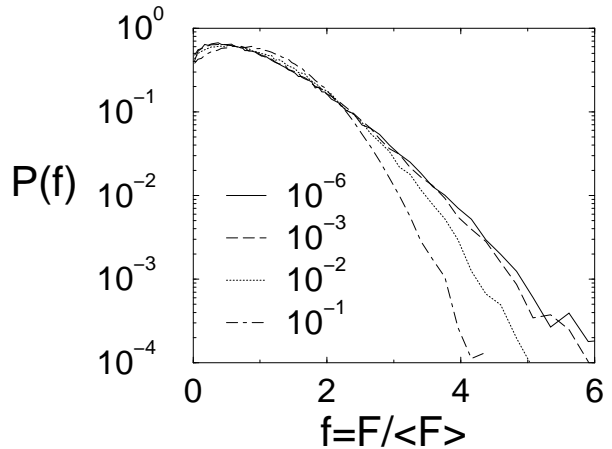}
  \caption{The distribution of normal contact forces, $P(F)$, for the purely
    repulsive, harmonic potential, at different compressions $\Delta \phi$.}
  \label{fig5}
\end{figure}

The results for $g(r<1)$ have implications for the distribution of
inter-particle normal forces, $P(F)$. This is shown in Fig.~\ref{fig5}. At
high compressions, $P(F)$ is well-described by a Gaussian, but the tail
straightens out towards an exponential as $\Delta \phi$ is lowered towards
zero. These results are consistent with previous results of Makse, et al.
\cite{makse1,makse2}, who studied sphere packings at fixed pressure. (As we
noted above, constant pressure corresponds to constant $\Delta \phi$.) The
Gaussian shape at high $\Delta \phi$ is consistent with expectations for
equilibrium systems interacting with a harmonic potential \cite{ohern1}.
However, these systems are at zero temperature, and it is unclear whether they
can be described by a nonzero effective temperature. The exponential behavior
at small $\Delta \phi$ agrees with experimental and simulation data on static
granular packings of hard particles, which necessarily exist at packing
fractions near $\phi_{c}$ \cite{mueth1,makse1,leo11}.

\begin{figure}[h]
  \includegraphics[width=7cm]{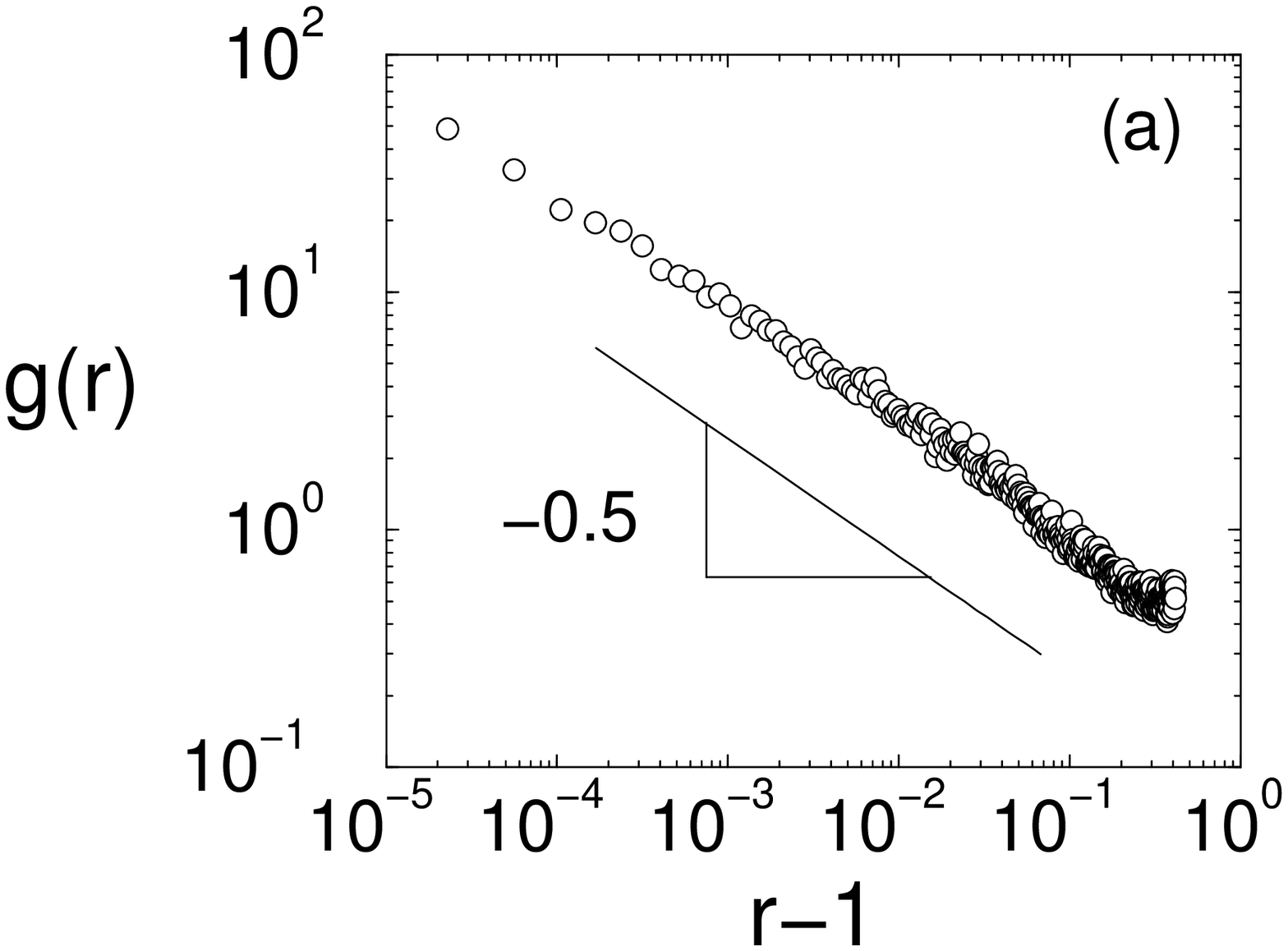}\\
  \includegraphics[width=7cm]{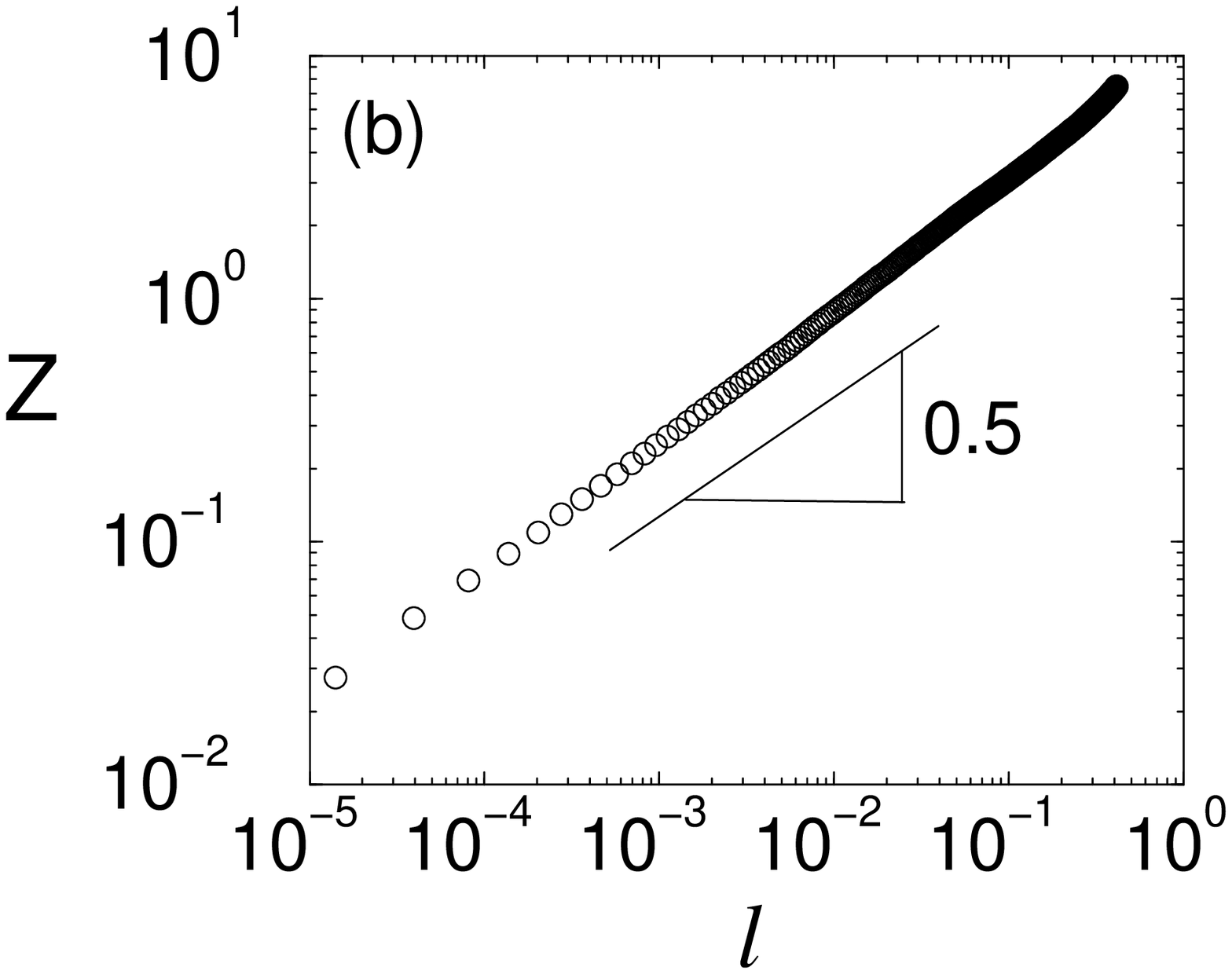}\hfil
  \caption{Behavior just above the first-neighbor peak for a system at $\Delta \phi
    =10^{-8}$. (a) $g(r>1)$ versus $r-1$. A power law with exponent of $-0.5$
    is indicated. (b) $Z(\ell)$ vs. $\ell \equiv r-1$, computed by numerically
    integrating $g(r) > 1$. A power law with exponent of $0.5$ is indicated.}
  \label{fig6}
\end{figure}

There is interesting behavior above the asymmetric first peak in $g(r)$ as
well as below it. Fig.~\ref{fig6}(a) shows that $g(r>1)$ versus $r-1$ varies
as a power-law for a system just above the transition at $\Delta\phi =
10^{-8}$:
\begin{equation}
  g(r>1) \propto [r-1]^{-\eta}
  \label{eqn4}
\end{equation}
with $\eta = 0.48\pm 0.03$. This result was first reported for
gravity-sedimented, granular packings \cite{leo9}, but over a much smaller
range in $g(r)$ than presented here. We note that there is a very slight knee
that occurs near $r-1 =3 \times 10^{-2}$. The asymptotic power law behavior
near $r=1$ should be determined only from the region below this knee. As we
will show below, this knee becomes more pronounced as $\Delta \phi$ increases.

The number of neighbors, $Z(\ell)$, that are separated by a distance of at
most $\ell$ \cite{finney3} is given by the integral
\begin{equation}
  Z(\ell) = 24\phi\int_{0}^{\ell} g(r^\prime) r^{\prime^{2}} dr^{\prime}.
\label{eqn5}
\end{equation}
This is shown in Fig.~\ref{fig6}(b). Therefore, Fig.~\ref{fig6}(a) and
Eq.~\ref{eqn4} imply, that for a system at the transition, $Z$ should increase
with $\ell$ as
\begin{equation}
  Z-Z_{\rm contact} \sim \ell^{1-\eta=0.52 \pm 0.03}
  \label{eqn6}
\end{equation}
where $\eta$ is defined in Eq.~\ref{eqn4} and $Z_{\rm contact} = 5.88$ is the
average number of neighbors per particle at contact at the transition. This
scaling is consistent with the one reported by O'Hern {\em et al.}
\cite{ohern2}, who looked at how the excess coordination number increased as a
system was incrementally compressed above $\phi_{c}$. We note that we have
found a similar exponent of $0.50 \pm 0.03$ using the Hertzian interaction
potential, $\alpha=2.5$, as we found for the harmonic potential, $\alpha=2$.

Donev et al. \cite{torquato5} also calculated the number of neighbors
$Z(\ell)$ near the transition, but for a system of hard spheres. When they
plotted $\log (Z(\ell)-Z_{c})$ versus $\log \ell$ they found that the slope
was closer to 0.6 than 0.5. They argued that this implies that $g(r>1) \propto
(1-r)^{-0.4}$ for hard spheres, which is different from our result for soft
repulsive spheres (Eq.~\ref{eqn6}). Based on this difference, they concluded
that the power-law behavior seen for $r > r_{\rm peak}$ is not universal and
depends on the interparticle potential as well as perhaps the procedure
leading to the $T=0$ configuration.

However, that conclusion does not follow from their analysis. The confusion
occurs because their results were for a system in which all the rattlers (that
is, particles without any neighbors) had been removed.  It is not surprising
that this changes the nature of the pair correlation function from a system,
such as ours, where {\it all} particles are considered.  That is, we analyze
the system directly as produced by energy minimization without performing the
additional step of removing particles that do not happen to be part of the
backbone of the structure.  Whether one chooses to remove rattlers or not
depends on the physics one wants to study.  Here we remark that it is more
natural to include rattlers in $g(r)$ if one is going to compare to any
experiment.  We also note that rattlers can join the backbone when the system
is perturbed, and can therefore influence the system’s response and stability,
as argued in Ref.~\cite{wyart2}. Finally, we claim that the conclusion of
Donev, et al.~that the power-law exponent depends on the potential of
interaction or the algorithm for creating the states (aside from removing one
class of particles by hand) is unfounded. Rather, the difference arises from
whether one studies systems with rattlers included or removed. When Donev, et
al. include rattlers in their analysis, they also find an exponent closer to
0.5 \cite{torquato4}, consistent with what we have found here and in previous
work \cite{leo9,ohern2,ohern3}.

There is a difference between $Z_{c}$ and $Z_{\rm contact}$.  $Z_{c}$ is the
number of contacts {\it only} for a system that has all rattlers removed from
the system.  If the rattlers are not removed from the system then $Z_{\rm
  contact} \leq Z_{c}$.  Because $g(r)$, for example measured in an
experiment, does not distinguish between rattlers and particles belonging to
the connected backbone of the structure, one must use $Z_{\rm contact} \approx
5.88 $ in Eqs.~\ref{eqn6}. Indeed, calculating $Z(\ell)$ either directly as in
Ref.~\cite{torquato5}, or integrating $g(r)$ via Eq.~\ref{eqn5}, are
identical only when using $Z_{\rm contact}=5.88$, and {\it not} $Z_{c}=6$ in
Eq.~\ref{eqn6}. If $N$ denotes the total number of particles, $N_{\rm
  overlaps}$ the number of overlapping pairs, and $N_{f}$ the total number of
rattlers, then $Z_{\rm contact}=2N_{\rm overlaps}/N$ while $Z_{c} = 2N_{\rm
  overlaps}/(N-N_{f})$.  Although only $2\%$ of the particles are rattlers at
the transition, their exclusion can produce a change of the power law
exponent.

\begin{figure}[!]
  \includegraphics[width=7cm]{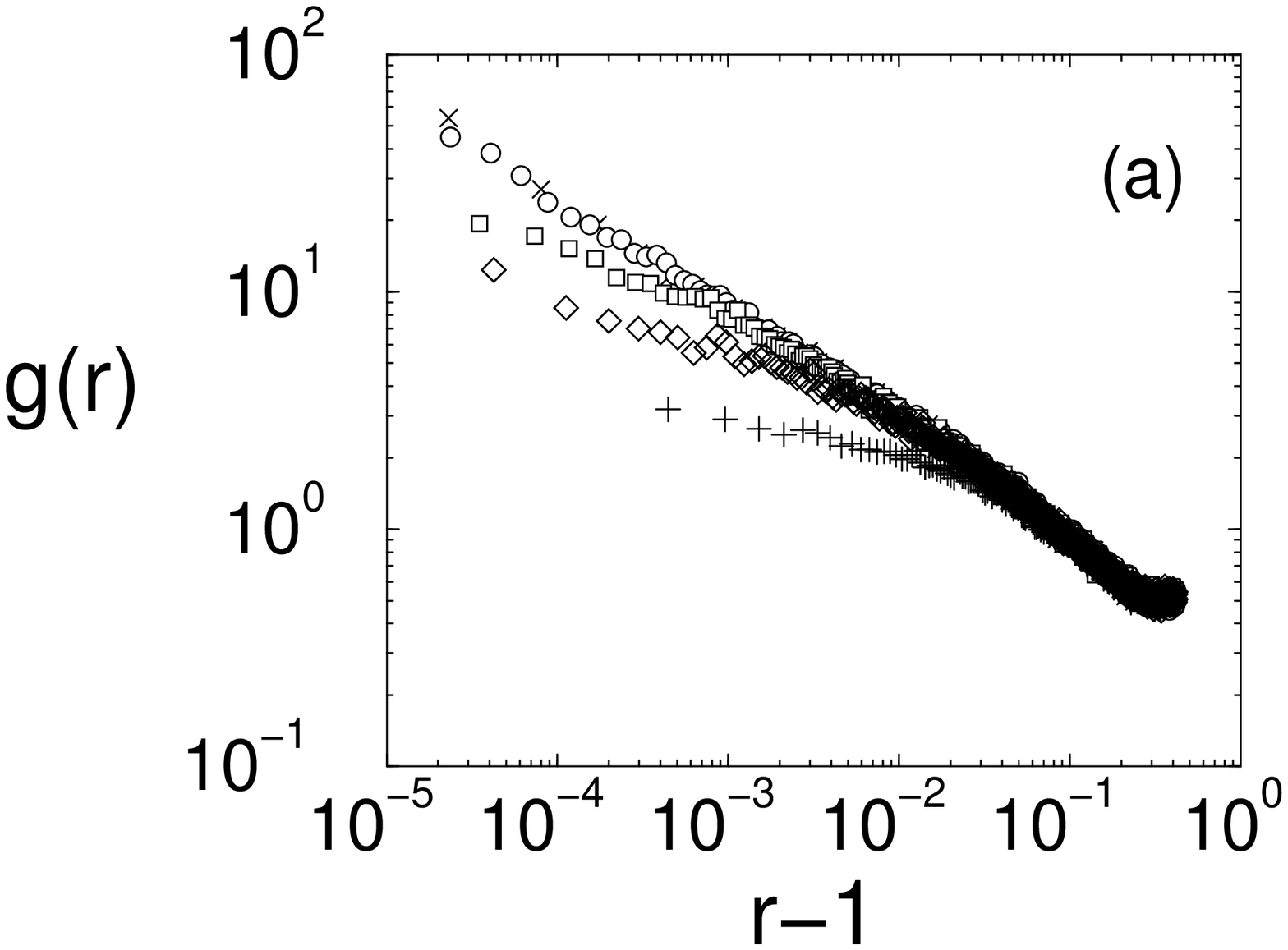}
  \includegraphics[width=7cm]{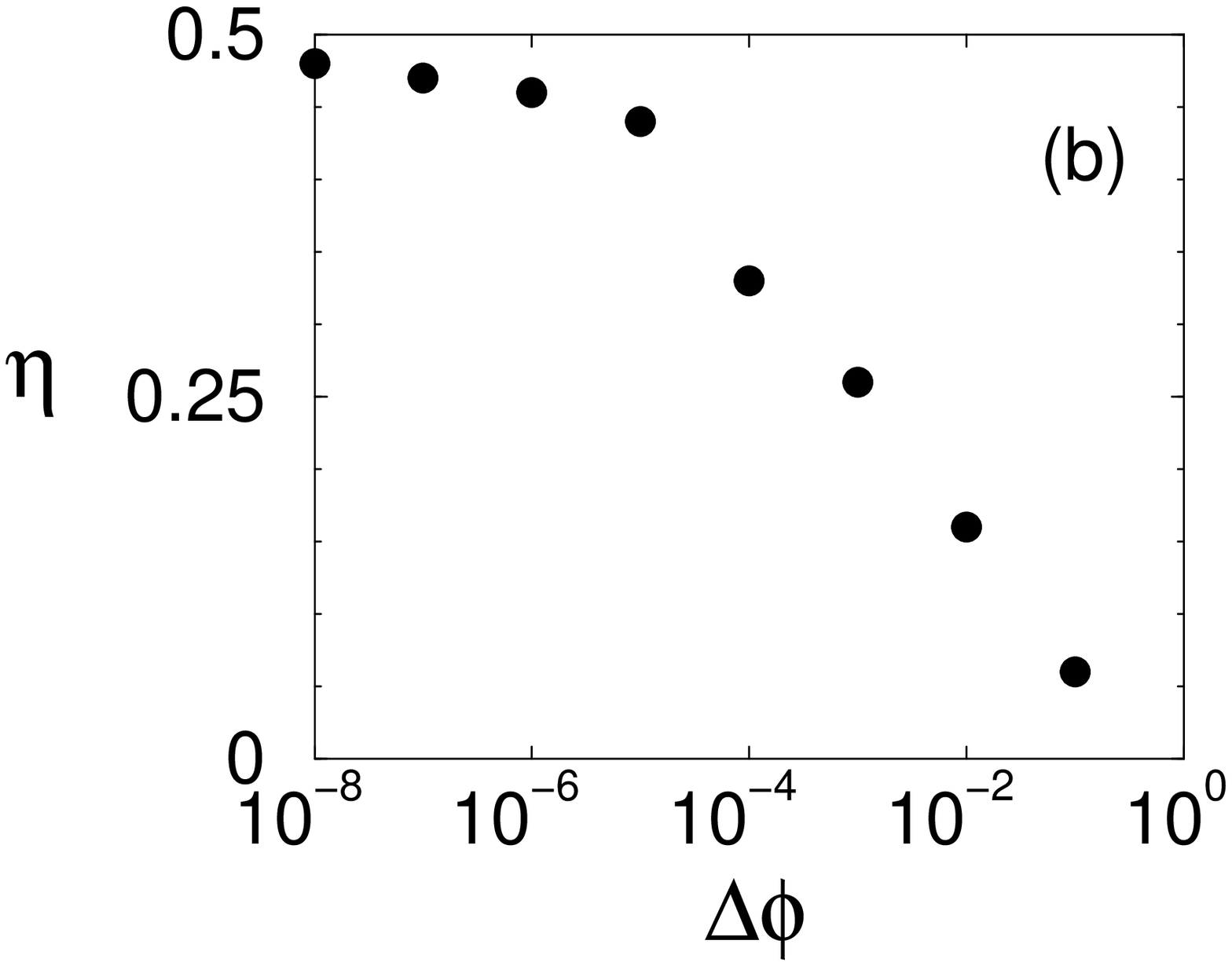}
  \includegraphics[width=7cm]{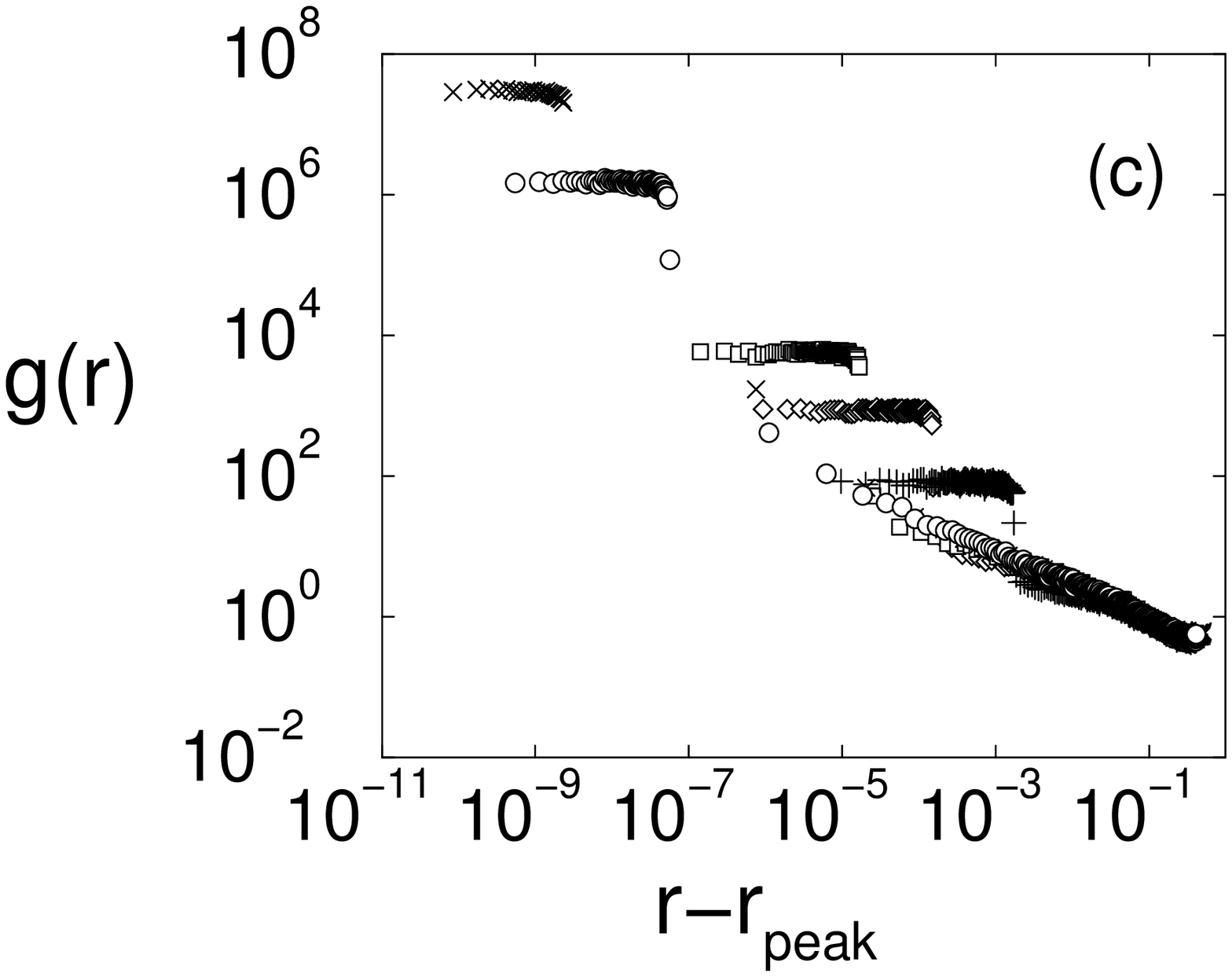}
  \caption{Pair distribution function $g(r)$, just above the first peak for $\Delta
    \phi = 10^{-8}$ ($\times$), $10^{-6}$ (circle), $10^{-4}$ (squares),
    $10^{-3}$~(diamonds), $10^{-2}$ ($+$). (a) $g(r)$ vs. $r-1$, showing the
    power-law behavior in $(r-1)$. (b) The exponent $\eta$ characterizing the
    power-law behavior of (a), as a function of $\Delta \phi$. (c) $g(r)$ from
    the nearest-neighbor, first peak at $r_{\rm peak}$ out to the first
    minimum at $r\approx 1.4$ as a function of $r-r_{\rm peak}$.}
  \label{fig7}
\end{figure}

We find that the observed power-law for $g(r)$ depends on $\Delta \phi$ (see
Fig.~\ref{fig7}(a)). As $\Delta \phi$ increases, the knee in $g(r)$ near $r-1
\approx 3 \times 10^{-2}$ becomes more prominent. In the region below this
knee ({\it i.~e.}, at smaller $r$), the slope of $\log [g(r)]$ versus $\log
[r-1]$ decreases.  We show this slope, $\eta$, as a function of $\Delta \phi$
in Fig.~\ref{fig7}(b). As $\Delta \phi$ approaches zero, the value of $\eta$
increases and approaches $\eta=0.5$. A similar trend was noted in the X-ray
tomography experiments of Aste et al. \cite{aste2}, who measured $g(r)$ inside
large, $3D$ granular packings. Note that Donev, et al. \cite{torquato5} study
configurations {\it below} the unjamming transition, and it is not clear if
the apparent value of $\eta$ also changes as the density is decreased below
$\phi_c$.
Fig.~\ref{fig7}(c) shows, for different values of $\Delta\phi$, how $g(r)$
behaves in the region from $r=r_{\rm peak}$ out to the first minimum at $r
\approx 1.4$. There are several notable features in $g(r)$ in the vicinity of
contact that are apparent when $g(r)$ versus $(r-r_{\rm peak})$ is plotted on
$\log-\log$ axes. There is a drop in $g(r)$ that occurs at $r=1$ for each
value of $\Delta \phi$ (note that this corresponds to different values of
$r-r_{\rm peak}$ for different $\Delta \phi$). The separation $r=1$
distinguishes particles that are overlapping from those which are just out of
contact. The magnitude of the jump decreases, and the extent of the power-law
region described by Eq.~\ref{eqn4} also decreases, as the system is
progressively compressed above the unjamming transition. The region beyond
contact is {\it relatively} unaffected by compression. (Although the power law
exponent changes slowly with $\Delta\phi$ as highlighted in
Fig.~\ref{fig7}(b), by far the largest change occurs in crossing from $r<1$ to
$r>1$.) This indicates that as the system is compressed, particles are
depleted from the region beyond contact, $r>1$, and are absorbed into the
contact region, $r<1$.

\begin{figure}[h]
  \includegraphics[width=7cm]{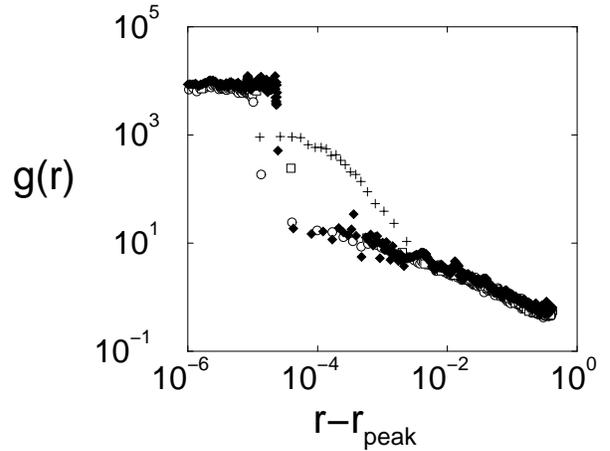}
  \caption{Pair distribution function $g(r)$, from the nearest-neighbor, first peak at
    $r_{\rm peak}$ out to the first minimum at $r\approx 1.4$, at $\Delta\phi
    = 10^{-4}$.  Different symbols represent different configuration protocols
    with and without temperature: conjugate gradient minimization to $T=0$
    (open circles), quenched molecular dynamics at $T=0$ (open squares),
    molecular dynamics at a very low temperature (solid diamonds), molecular
    dynamics at a higher temperature (+). The jump in $g(r)$ persists to small
    but nonzero temperature, and is smoothed out at high enough temperatures.}
  \label{fig8}
\end{figure}

We have determined that this drop in $g(r)$ is not an artifact of the
zero-temperature system. We have used several different protocols for creating
particle configurations, as shown in Fig.~\ref{fig8} and described in its
caption.  Evidently, the drop in $g(r)$ persists to finite temperature.

Why should there be an abrupt drop in $g(r)$ at $r=1$? One possibility is
that, to a first approximation, compression only changes the nearest-neighbor
spacing of particles that are already in contact, that is, within a distance
$r<1$ of each other. Particles separated by a distance $r>1$ may not be moved
appreciably closer to one another by compression. Instead, we suggest that
upon compression, the movement of particles that are not yet overlapping is
predominantly in a direction perpendicular to the line connecting them.  This
is consistent with our data.

\begin{figure}[h]
  \includegraphics[width=7cm]{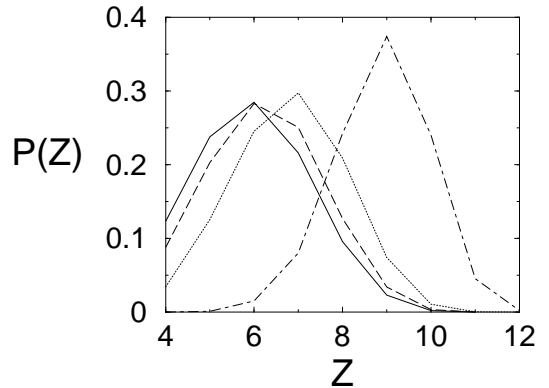}
  \caption{The distribution of the number of overlapping neighbors per particle, $Z$,
    at several different compressions. Key same as Fig.~\ref{fig5}.}
  \label{fig9}
\end{figure}

As the system is compressed and particles are incorporated into the contact
region, the number of overlapping pairs increases.  As mentioned above, the
average number of overlapping neighbors per particle, $Z$, increases with
compression as $Z-Z_{\rm contact} \propto \Delta \phi^{0.5}$. As this occurs,
the distribution of $Z$ values, $P(Z)$, also shifts. This is shown in
Fig.~\ref{fig9}. Close to the transition, most particles have 6 overlapping
neighbors. As the system is compressed to $\Delta \phi = 0.1$, the maximum
shifts to $Z=9$ but there is still no observable weight at $Z=12$.

\section{Split second peak of $g(r)$}
\begin{figure}[h]
  \includegraphics[width=8cm]{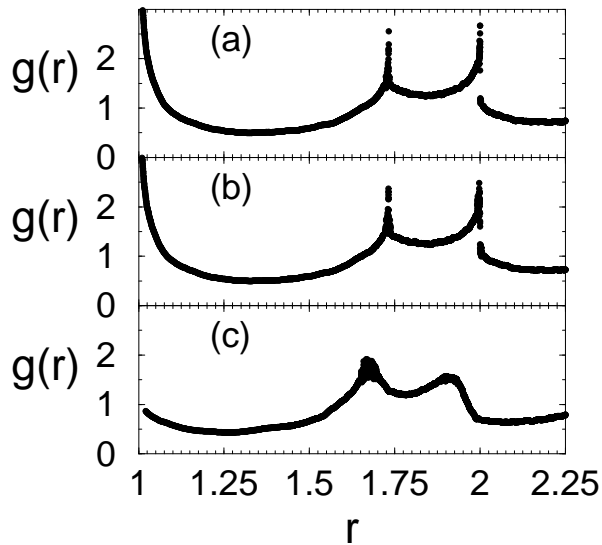}
  \caption{$g(r)$ region around the split second peak for $\Delta\phi =$
    (a) $10^{-6}$, (b) $10^{-3}$, and (c) $10^{-1}$.}
  \label{fig10}
\end{figure}

The approach of the unjamming point is also apparent in features associated
with the second peak in $g(r)$. Figure \ref{fig10} shows this region of $g(r)$
for several values of $\Delta\phi$. For the system closest to jamming, at
$\Delta\phi = 1\times 10^{-6}$, (Fig.~\ref{fig10}(a)) there is a pronounced
splitting of the second peak in $g(r)$ into two sub-peaks, located at
$r=\sqrt{3}$ and $r=2$.  Such a splitting of the second peak has long been
known \cite{cargill1}. Indeed, the emergence of a split second peak was one of
the early criteria used to signal the onset of the glass phase in supercooled
liquids \cite{abraham1}.

\begin{figure}[h]
  \includegraphics[width=4cm]{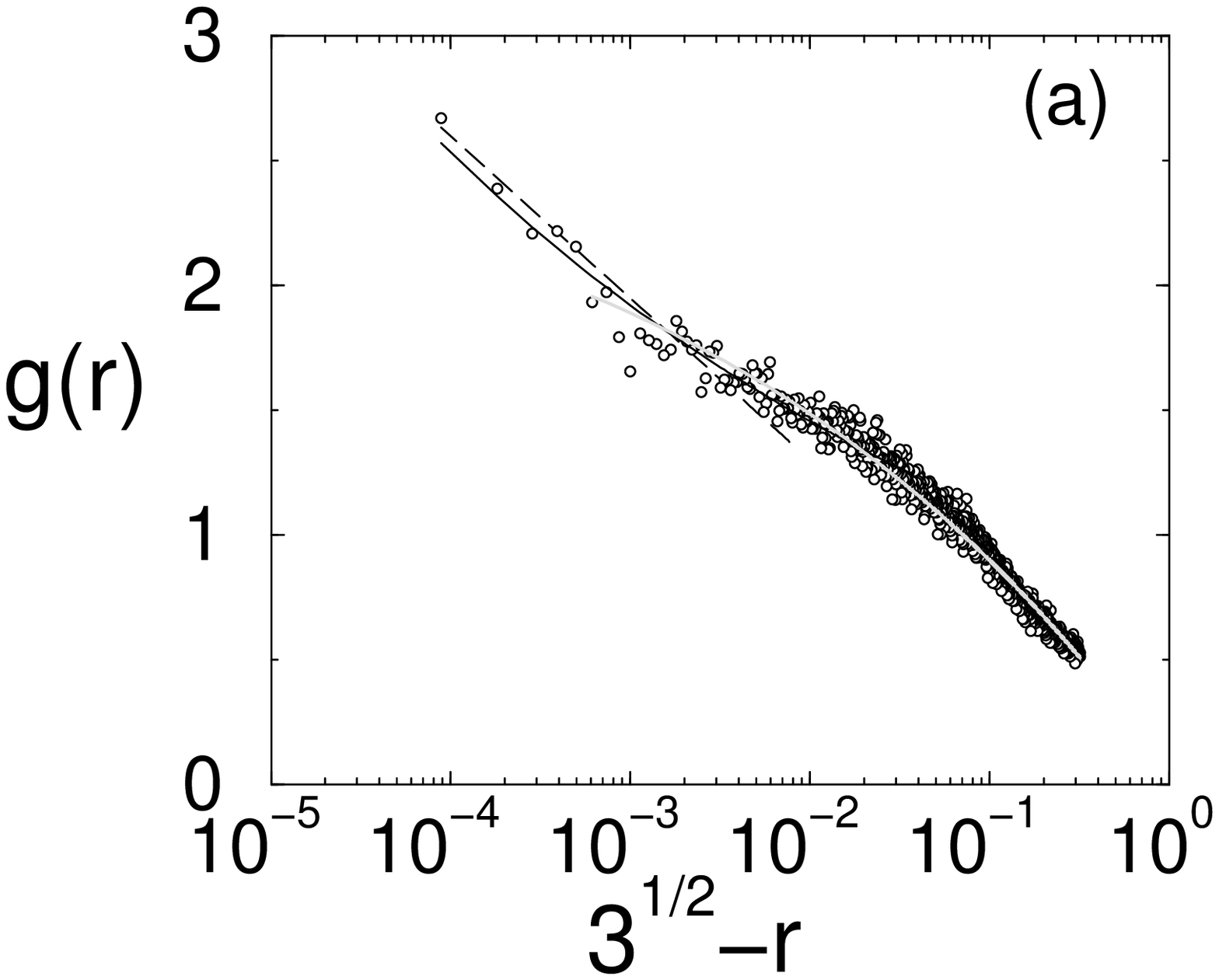}
  \includegraphics[width=4cm]{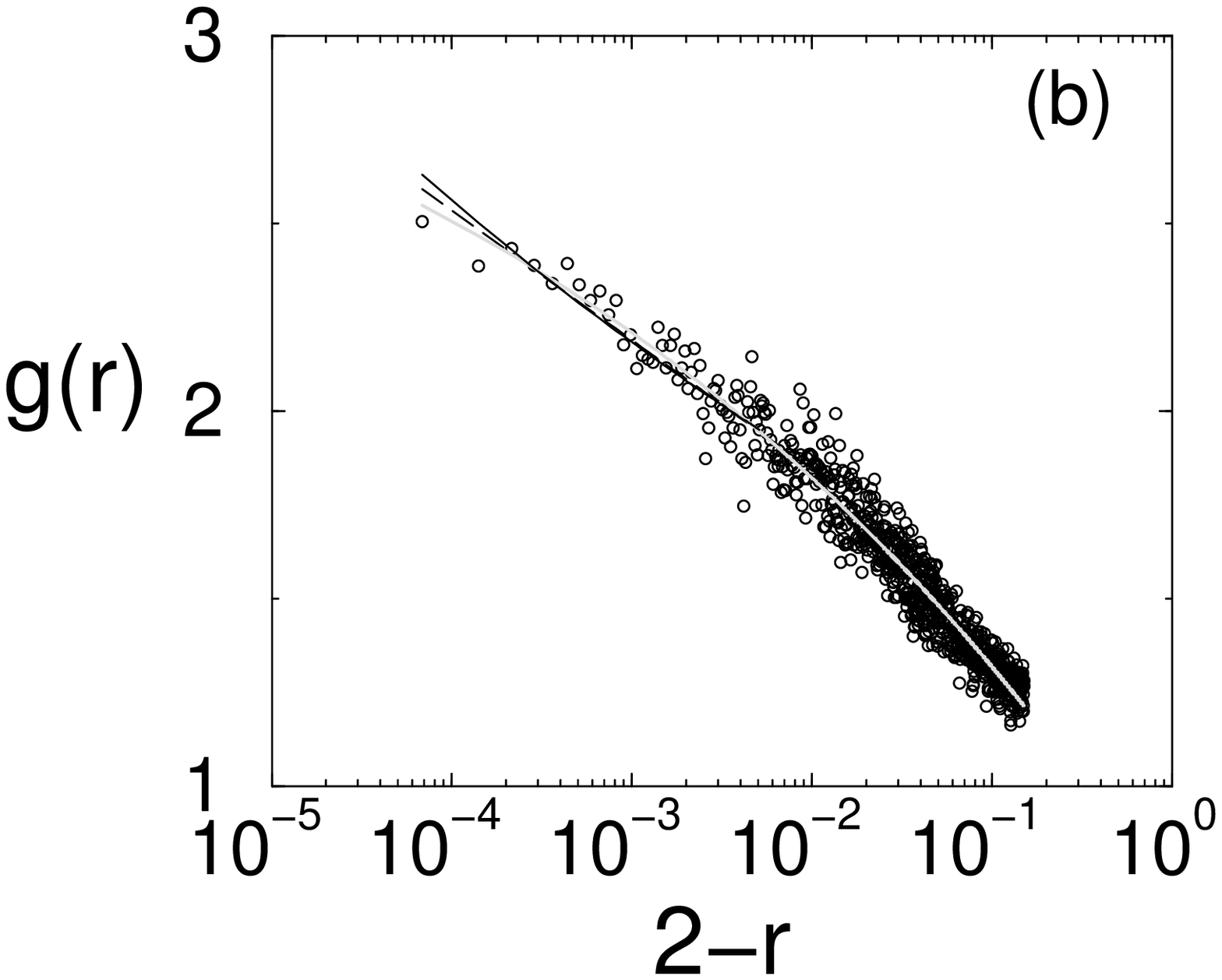}
  \caption{Functional form of the sub-peaks of the split-second peak in $g(r)$ for
    $\Delta\phi = 10^{-6}$. Left-hand sub-peak at $r=\sqrt{3}$.  Right-hand
    sub-peak at $r=2$. The solid line corresponds to a power-law fit, the
    dashed line to a logarithmic fit, and the grey line to a shifted power-law
    fit.}
  \label{fig11}
\end{figure}

On the small-$r$ side of both of these sub-peaks we find that $g(r)$ increases
rapidly. In Fig.~\ref{fig11}, we have attempted to characterize these features
for the system closest to the transition.  In that figure we plot $g(r)$
versus $\log(\sqrt{3} -r)$ and $\log(2-r)$. We have not been able to tell
whether $g(r)$ itself diverges or whether the divergence appears only in its
slope, $dg(r)/dr$. We would need to average over many more systems to tell
these two possibilities apart unambiguously. Our best fits to the two possible
cases are:
\begin{eqnarray}
  g(r) & = & a_{1} (\sqrt{3} - r)^{-a_{2}}\\
\label{eqn7}
  g(r) & = & g(\sqrt{3}) - b_{1} (\sqrt{3} - r)^{b_{2}} \\
\label{eqn8}
  g(r) & = & -c_{1} \log(\sqrt{3}-r)  
\label{eqn9}
\end{eqnarray}
and
\begin{eqnarray}
  g(r) & = & d_{1} (2 -r)^{-d_{2}}\\
\label{eqn10}
  g(r) & = & g( 2) - e_{1} (2-r)^{e_{2}}\\
\label{eqn11}
  g(r) & = & -f_{1} \log (2-r) 
\label{eqn12}
\end{eqnarray}
where the fit parameters are provided in Table ~\ref{table1}.
\begin{table}[h]
  \caption{Fit parameters for the two sub-peaks, situated at $r=\sqrt{3}$ and $r=2$, respectively, that make up the split second peak in $g(r)$. For $\Delta\phi=10^{-6}$.}
\begin{tabular}{c|c}[h]
First sub-peak & Second sub-peak\\ \hline
$a_{1}$ = 0.9 & $d_{1}$ = 1.1 \\
$a_{2}$ = 0.12 & $d_{2}$ = 0.08 \\ \hline
$b_{1}$ = 2.75 & $e_{1}$ = 2.8 \\
$b_{2}$ = 0.16 & $e_{2}$ = 0.12 \\ \hline
$c_{1}$ = 0.25 & $f_{1}$ = 0.15 \\ \hline
\end{tabular}
\label{table1}
\end{table}

On the high-$r$ side of each sub-peak, it is clear from Fig.~\ref{fig10}(a)
that there is both a step-function drop-off \cite{finney1,torquato5} and an
additional smooth decrease of $g(r)$ with increasing $r$. Although, from these
data, we cannot unambiguously determine a functional form for this last smooth
decrease, the data suggest that it might be fit with
\begin{equation}
  g(r)=a-(r-r_s)^{b},
\label{eqn13} 
\end{equation}
where $0<b<1$ and $r_s=\sqrt{3}$ and $2$ for the first and second sub-peaks,
respectively.  As the system is compressed above the transition, the structure
around both sub-peaks becomes rounded.  In particular, it is only in the limit
of $\Delta\phi \rightarrow 0$ that the drop-off becomes a sharp step-function.

Here, we suggest a simple interpretation of the origin of the step-function
drop-off in $g(r)$ on the high-$r$ side of both sub-peaks.  At the jamming
transition, each particle must be held in place by particles that are just in
contact with no overlap. At that point there is an average of precisely $Z_{c}
= 6$ neighbors per particle in the force network.  It seems plausible that the
second-nearest-neighbor peaks originate from pairs of particles that have at
least one neighbor in common, while pairs that do not have a common neighbor
contribute only to a slowly-varying background.

\begin{figure}[h]
  \includegraphics[width=4cm]{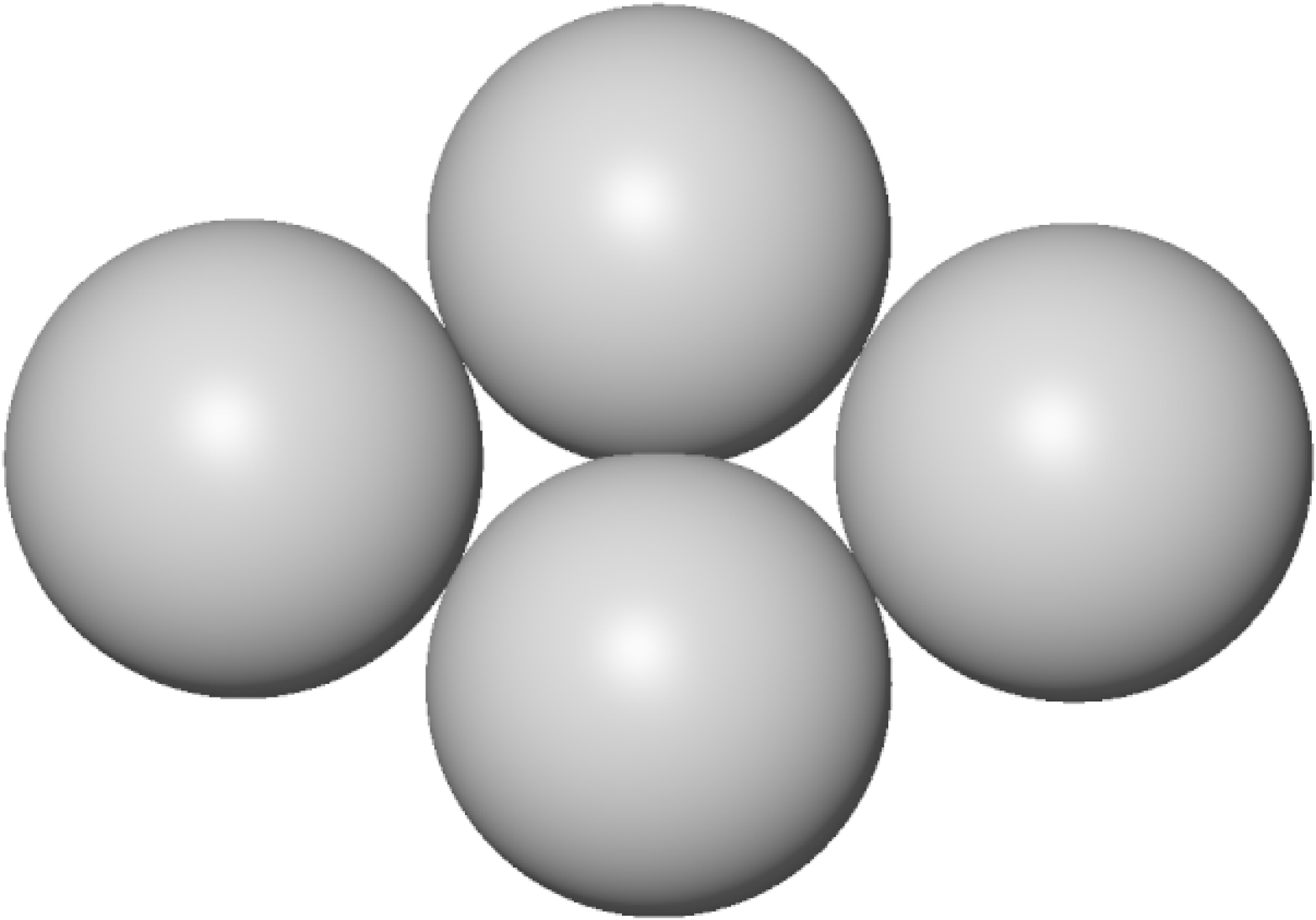}
  \includegraphics[width=4cm]{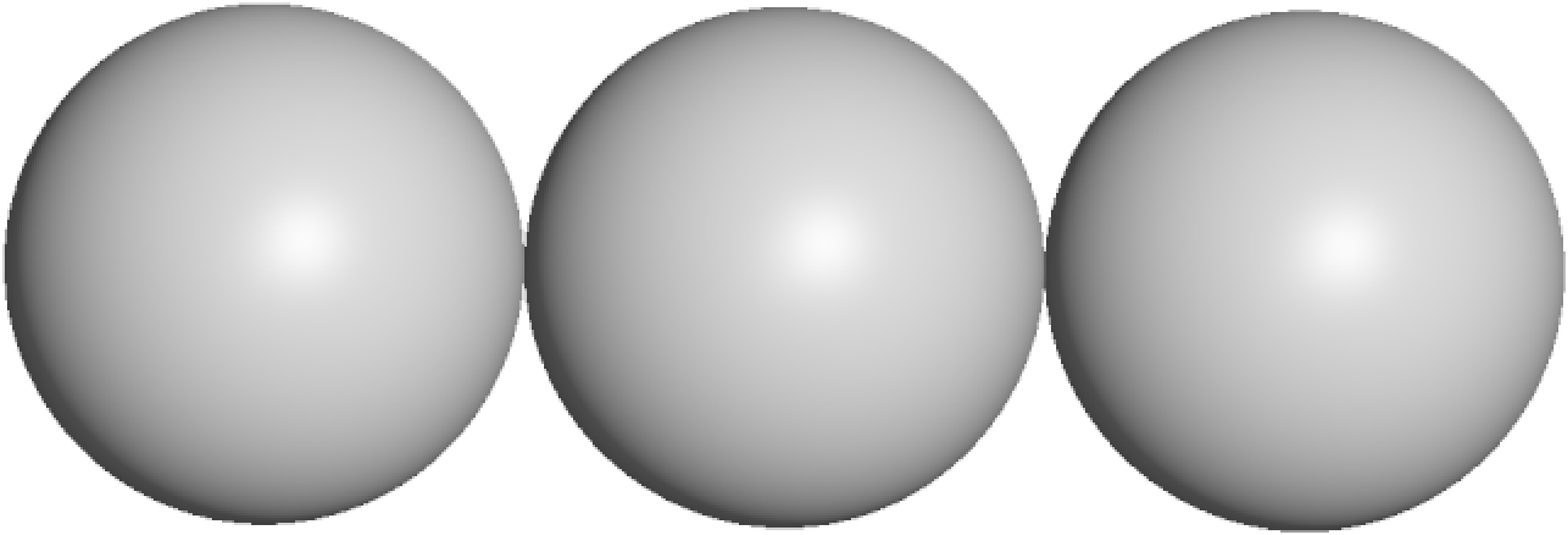}
  \caption{Left: The particles at the left and at the right share two
    common neighbors and are separated by a distance $r=\sqrt{3}$. Right: The
    particles at the left and the right share one common neighbor and are
    separated by $r=2$.}
  \label{fig12}
\end{figure}

The separation $r=2$ corresponds to two particles on diametrically opposite
sides of a common neighbor (Fig.~\ref{fig12}, right figure). This is the
largest distance that can separate two particles that have {\it one} common
neighbor. The separation $r = \sqrt{3}$ corresponds to the largest possible
separation between two particles, $i$ and $j$, that have {\it two} common
neighbors (Fig.~\ref{fig12}, left figure). In this case the four particles lie
in the same plane; the two common neighbors touch and the particles $i$ and
$j$ are on opposite sides of the crease between them. When $r$ is only
slightly greater than $2$, there can no longer be any contribution from two
particles that share a common neighbor. Similarly, when $r$ is only slightly
greater than $\sqrt{3}$, there can be no contribution from pairs of particles
that share two common neighbors. This leads to step-function drop-offs at
$r=\sqrt{3}$ and $r=2$.

\begin{figure}[h]
  \includegraphics[width=8cm]{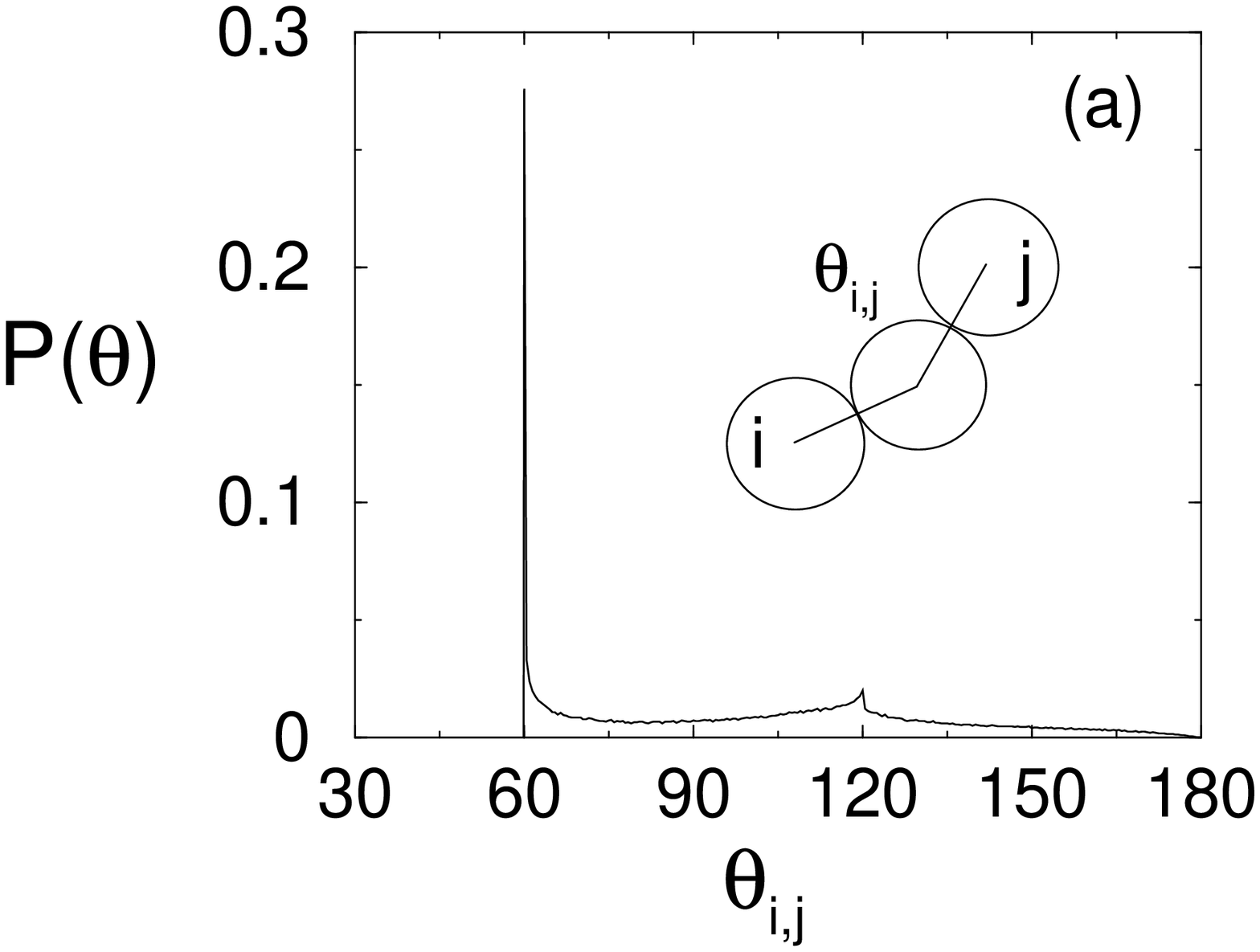}\\
  \includegraphics[width=3.5cm]{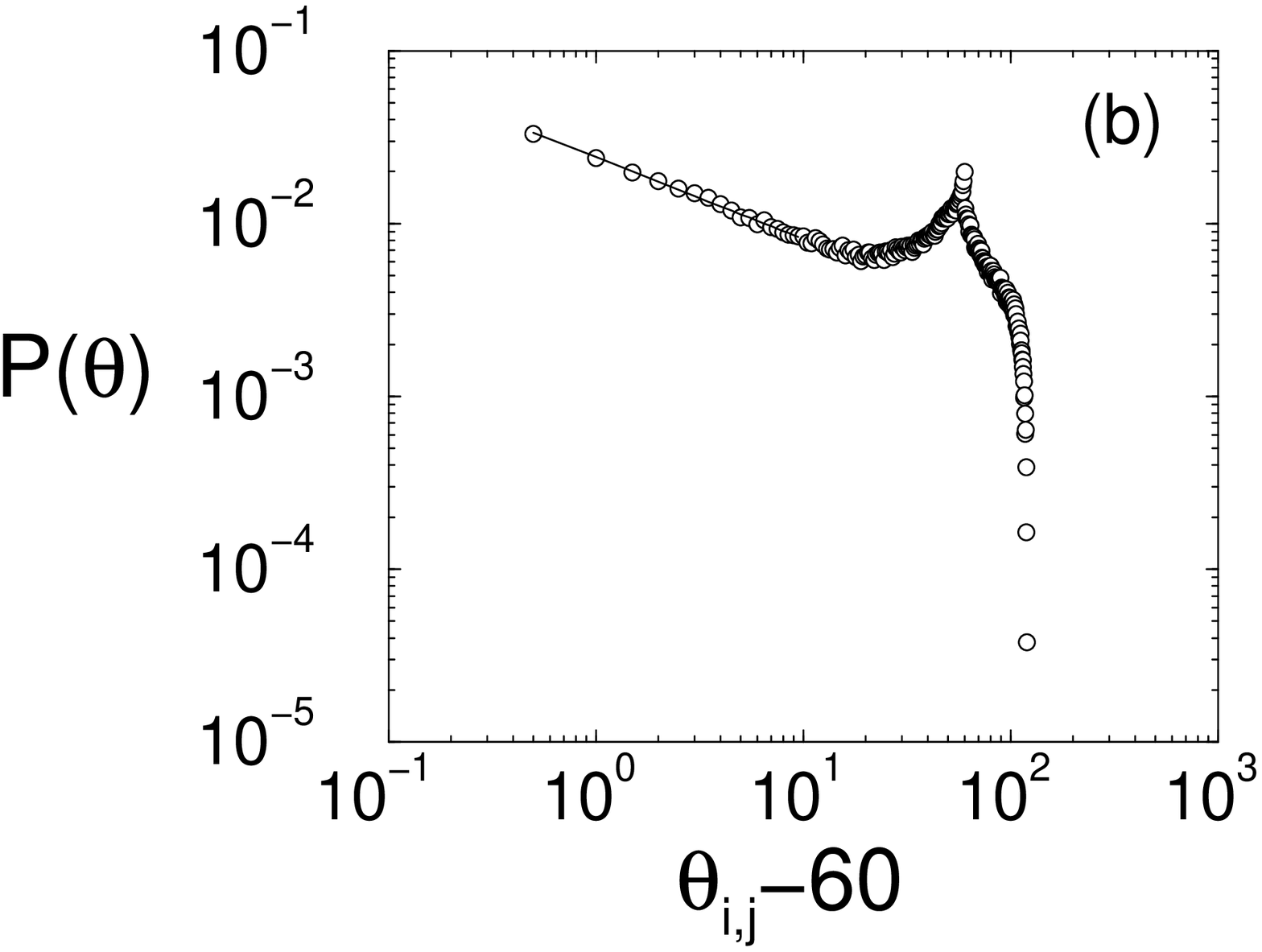}\hfil
  \includegraphics[width=3.5cm]{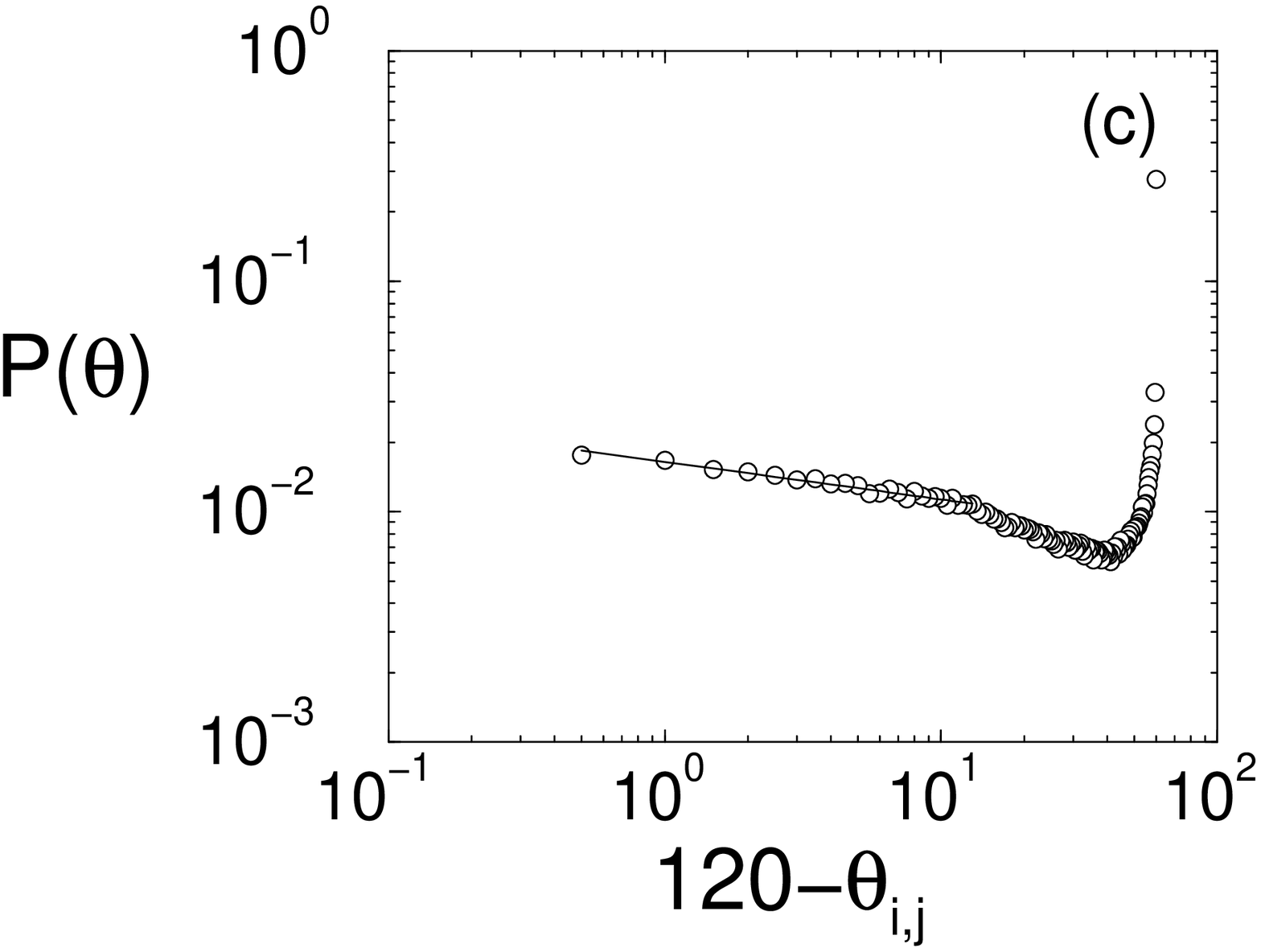}\hfil
  \includegraphics[width=3.5cm]{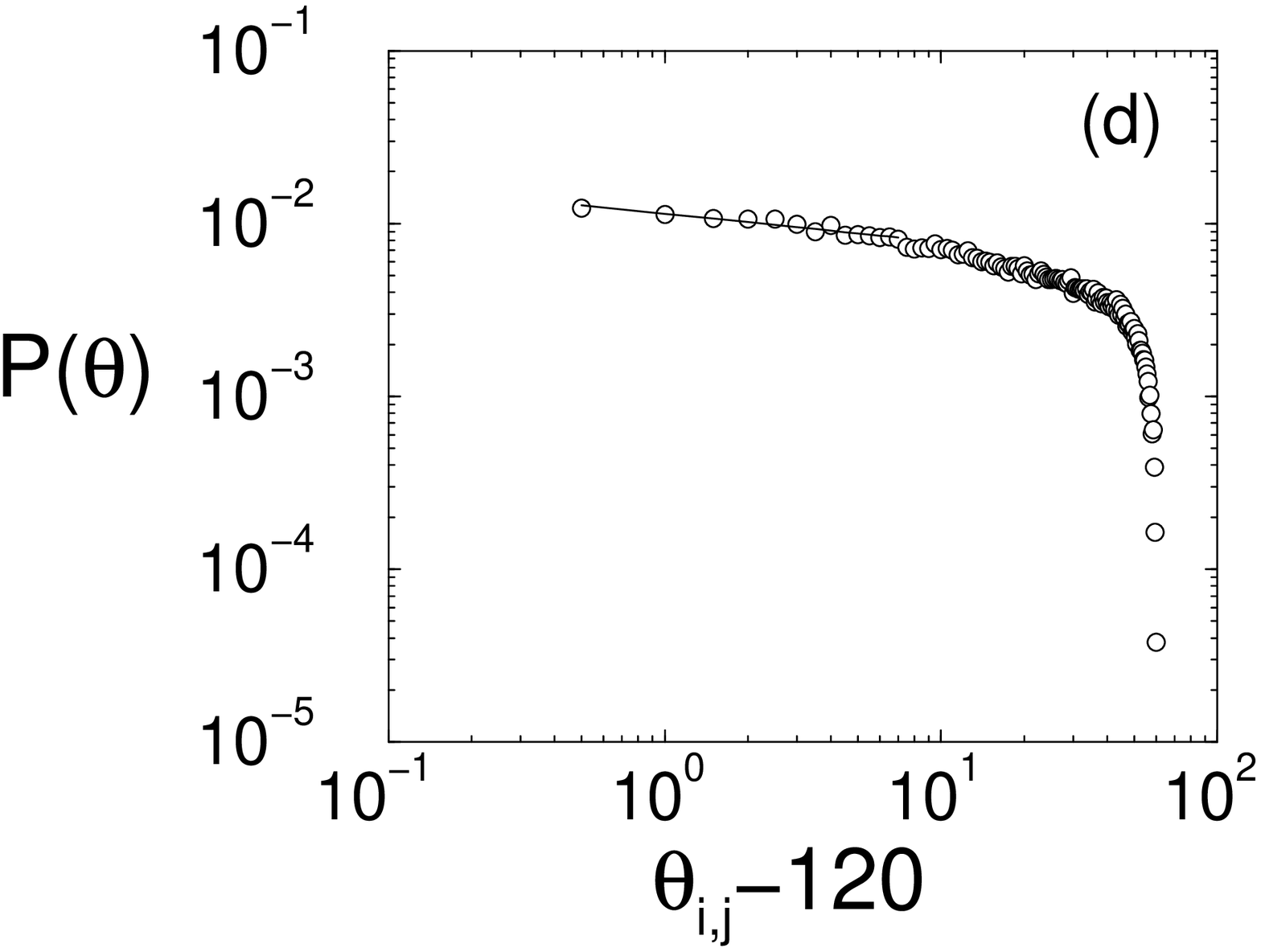}\hfil
  \includegraphics[width=3.5cm]{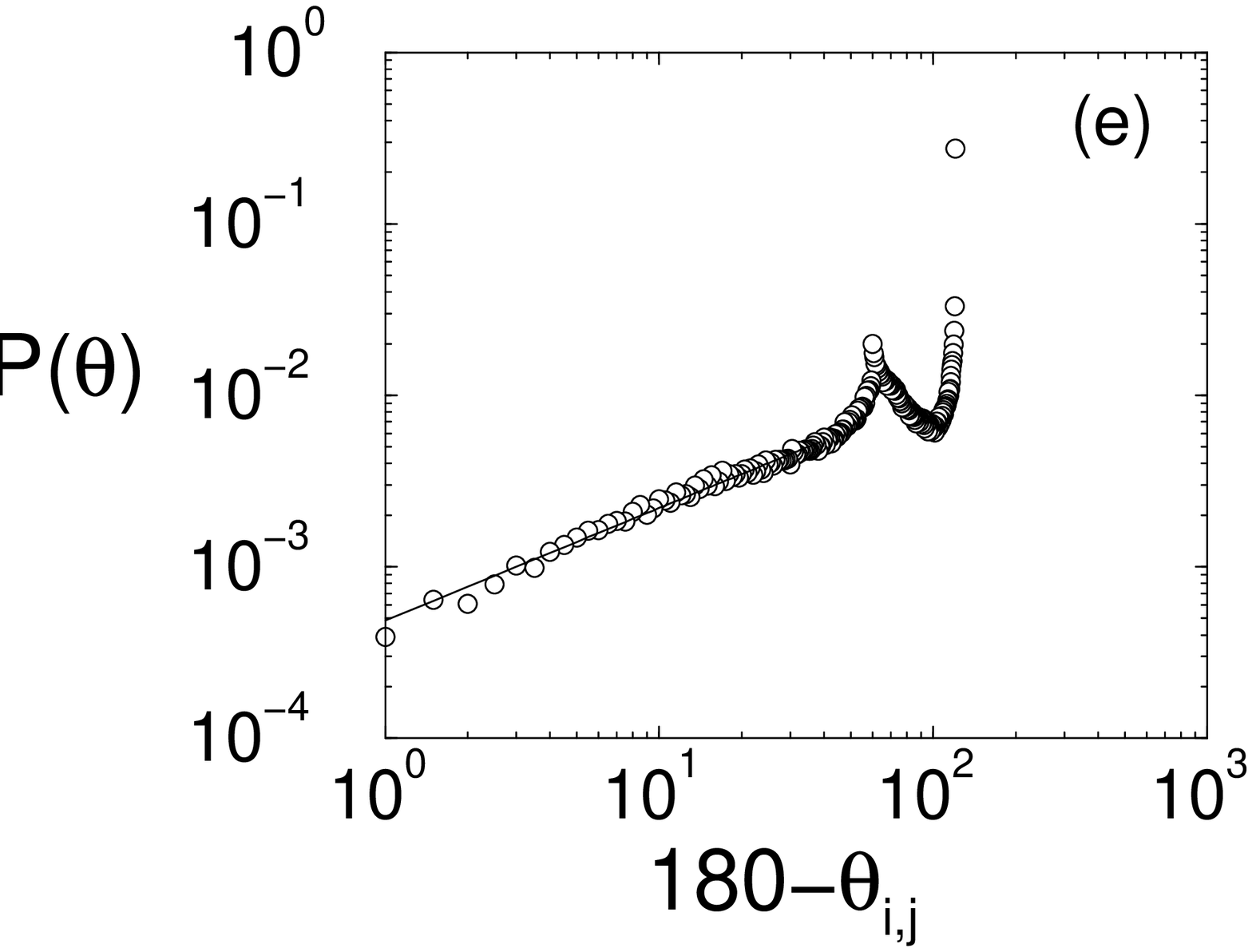}\hfil
  \caption{(a) The three-particle angular correlation, $P(\theta_{i,j})$, versus
    $\theta_{i,j}$ for $\Delta\phi=10^{-6}$. Bottom four panels are fits to
    the regions around; (b) $\theta=60^{\circ}$, (c)
    $\theta=120^{\circ}$-left, (d) $\theta=120^{\circ}$-right, (e)
    $\theta=180^{\circ}$. The power law exponents are $\approx$ 0.5, 0.17,
    0.17, and 0.75, respectively}
  \label{fig13}
\end{figure}

Support for the hypothesis that the drop-offs are due to pairs of particles
that share common neighbors can be obtained from the angular correlations
between two neighbors of a common particle.  We show this angular correlation,
$P(\theta)$, in Fig.~\ref{fig13}, where $\theta$ is defined in the sketch in
the inset. At $\theta = \pi$ (which corresponds to particles separated by
$r=2$) Fig.~\ref{fig13}(d) shows that the distribution goes to zero as,
\begin{equation}
  P(\theta) \sim (\pi-\theta)^{x=0.75}
\label{eqn14}
\end{equation}
If the sub-peak at $r=2$ arises from pairs that share one common neighbor,
then the form of $g(r)$ just below $r=2$ should be related to the form of
$P(\theta)$ just below $\theta=\pi$ by
\begin{equation}
  g(r) = P(\theta) d\theta/dr
\label{eqn15}
\end{equation}
where the Jacobian factor diverges as
\begin{equation}
  d\theta_{i,j}/dr = \frac{1}{\cos\frac{\theta_{i,j}}{2}} = \frac{2}{\sqrt{4-r^{2}}}
\label{eqn16}
\end{equation}
Thus, Eq.~\ref{eqn15} implies
\begin{equation}
  g(r) \sim (2-r)^{\frac{x-1}{2}=-0.12}
\label{eqn17}
\end{equation}
This is in reasonable agreement with the fit in Eq.~\ref{eqn10}, where we
found $g(r) \sim (2-r)^{-d_2 = -0.08}$, implying that the second sub-peak does
indeed arise from two particles that share a common neighbor and that this
leads to the observed step-function drop-off just above $r=2$. However, we
should note that this argument does not necessarily imply that $g(r)$ diverges
at $r=2$, as indicated by Eq.~\ref{eqn17}, since we cannot determine the
behavior of $P(\theta)$ asymptotically close to $\pi$ with sufficient
accuracy.

By a similar argument, if the first sub-peak arises from pairs of particles
that share two common neighbors, then the peak at $r=\sqrt{3}$ corresponds to
$\theta=2\pi/3$.  For $\theta<2\pi/3$, we find
\begin{equation}
  P(\theta) \sim (2\pi/3-\theta)^{-0.17}
\label{eqn18}
\end{equation}
This behavior implies $g(r) \sim (\sqrt{3}-r)^{-0.17}$; this result is not too
different from the fit in Eq.~7, which suggests $g(r) \sim
(\sqrt{3}-r)^{-0.12}$.  This consistency check suggests that the first
sub-peak does indeed arise from pairs of particles that share two common
neighbors.

Although pairs with two common neighbors cannot contribute to $g(r)$ for
$r>\sqrt{3}$, pairs with only one common neighbor can still contribute.  Such
pairs may account for the smooth decrease in $g(r)$ (see Eq.~\ref{eqn13}) just
above $r=\sqrt{3}$.

\section{Discussion}
We have shown that the unjamming transition is accompanied by several features
in the pair correlation function:
\begin{itemize}

\item 

  A delta-function at $r=1$ with area $Z_{\rm contact}<6$.

\item 

  A power-law at $r>1$ of the form $g(r) \sim (r-1)^{-0.5}$.

\item

  A sub-peak at $r=\sqrt{3}$ that either diverges or has diverging
  slope as $r \rightarrow \sqrt{3}^-$ and that has a step-function
  drop-off just above $r=\sqrt{3}$.

\item
  
  A sub-peak at $r=2$ that either diverges or has diverging slope as $r
  \rightarrow 2^{-}$ and that has a step-function drop-off just above $r=2$.

\end{itemize}
These features appear for both harmonic and Hertzian repulsions, and therefore
seem to be purely geometrical features of the jamming/unjamming transition at
zero temperature.

Here, we suggest that some of these structural features are reflected in a
less extreme form in the two primary empirical criteria that have been used
extensively in the literature to identify the glass transition.  The first is
the ratio, $\cal{R}$, of the first minimum to the first maximum in $g(r)$.  As
the temperature $T$ of a glass-forming liquid decreases, ${\cal R}$ decreases.
Wendt and Abraham \cite{abraham1} proposed that ${\cal{R}}=0.14$ corresponds
to a reasonable estimate of the glass transition temperature, $T_{g}$.  We
note that at the unjamming transition we have studied here, ${\cal R}=0$.  It
is possible that the decrease in ${\cal R}$ observed as $T$ is lowered towards
$T_{g}$ is a remnant of the vanishing of ${\cal R}$ that occurs at the
unjamming transition.  The latter property may be the underlying reason for the
success of this empirical criterion.

A second popular empirical criterion concerns the second peak of $g(r)$.  As
$T$ is lowered towards $T_g$, the second peak splits into two sub-peaks. The
flattening of the second peak that signals its splitting has been used to
identify $T_{g}$ \cite{abraham1}. We suggest that the splitting may reflect
the singular sub-peaks that occur at jamming/unjamming transition.  The
singular nature of the splitting at this transition may provide the
fundamental underpinnings of this criterion.

\acknowledgments 
We thank Bulbul Chakraborty and Corey O'Hern for instructive discussions and
Salvatore Torquato for comments on our manuscript.  We also gratefully
acknowledge the support of NSF-DMR-0087349 (AJL), NSF-DMR-0352777 (SRN),
DE-FG02-03ER46087 AJL,LES), and DE-FG02-03ER46088 (SRN,LES).

\end{document}